\documentclass[english]{article}
\usepackage[T1]{fontenc}
\usepackage[latin9]{inputenc}
\usepackage{amsmath}
\usepackage{amssymb}
\usepackage{stackrel}

\makeatletter

\providecommand{\tabularnewline}{\\}

\usepackage{jheppub}
\title{Quivers for 3-manifolds: the correspondence, BPS~states, and 3d $\mathcal{N}$=2 theories}
\author{Piotr Kucharski}
\affiliation{Walter Burke Institute for Theoretical Physics, California Institute of Technology, \\ 1200~E.~California Blvd. Pasadena, CA 91125, USA} 
\affiliation{Faculty of Physics, University of Warsaw,\\ ul. Pasteura 5, 02-093 Warsaw, Poland}
\emailAdd{piotrek@caltech.edu}
\newtheorem{conj}{Conjecture}
\abstract{
We introduce and explore the~relation
between quivers and 3-manifolds with the~topology of the~knot
complement. This idea can be viewed as
an~adaptation of the~knots-quivers
correspondence to Gukov-Manolescu invariants of knot
complements (also known as $F_K$ or~$\hat{Z}$). 
Apart from assigning quivers to complements
of $T^{(2,2p+1)}$ torus knots, we study the~physical 
interpretation in terms of the~BPS~spectrum and general
structure of 3d $\mathcal{N}=2$ theories associated to
both sides of the~correspondence. We also make a~step towards 
categorification by proposing a~$t$-deformation of 
all objects mentioned above.
}

\makeatother

\usepackage{babel}
\begin{document}
\maketitle

\section{Introduction}

In recent years many interesting relations between quantum field theory,
string theory, low-dimensional topology, and representation theory
have been studied. In this paper we focus on the~knots-quivers correspondence
\cite{KRSS1707short,KRSS1707long} and the~Gukov-Manolescu invariants
of knot complements \cite{GM1904}. (We will abbreviate them to GM~invariants
and denote $F_{K}$ -- often this symbol is also used as their name.)

The~core of the~knots-quivers correspondence is an~equality (after
appropriate change of variables) between the~generating series of
symmetrically coloured HOMFLY-PT polynomials of a~knot and a~certain
motivic generating series associated to the~quiver. This implies
that Labastida-Mari\~{n}o-Ooguri-Vafa (LMOV) invariants of knots
\cite{OV9912,LM01,LMV00} can be expressed in terms of motivic Donaldson-Thomas
(DT) invariants of associated quivers \cite{KS0811,KS1006}, which
are known to be integer \cite{Efi12}. In consequence, by assigning
a~quiver to the~knot, we automatically prove the~integrality of
LMOV invariants, which is the~statement of the~LMOV conjecture~\cite{OV9912,LM01,LMV00}.
This was done for all knots up to 6 crossings and infinite families
of $T^{(2,2p+1)}$ torus knots and twist knots in \cite{KRSS1707long}.
More systematic approach in terms of tangles provided a~proof of
the~knots-quivers correspondence for all two-bridge knots \cite{SW1711}
and recently for all arborescent knots \cite{SW2004}, however finding
a~general proof remains to be an~open problem.

Geometric and physical interpretation of the~knots-quivers correspondence
given in \cite{EKL1811,EKL1910} connects it with Gromov-Witten invariants
and counts of holomorphic curves, showing that the~quiver encodes
the~construction of BPS spectrum from the~basic states. This spectrum
is shared by dual 3d $\mathcal{N}=2$ theories associated to the~knot
complement and the~quiver, and respecitve counts of BPS states are
given by LMOV and DT invariants. The~moduli spaces of vacua of these
theories are encoded in the~graphs of $A$-polynomials \cite{AFGS1203,AV1204,FGS1205,GLL1604,FGSS1209}
and their quiver versions \cite{EKL1811,EKL1910,PSS1802,PS1811}.
Another recent results on the~knots-quivers correspondence include
a~relation to combinatorics of counting paths \cite{PSS1802}, a~consideration
of more general case of topological strings on various Calabi-Yau
manifolds \cite{PS1811}, and a~connection with the~topological
recursion \cite{LNPS2005}.

The origins of GM invariants lie in attempts of categorification of
the~Witten-Reshetikhin-Turaev (WRT) invariants of 3-manifolds. In
order to solve the~problem of non-integrality of WRT invariants Gukov,
Pei, Putrov, and Vafa introduced new invariants of 3-manifolds \cite{GPV1602,GPPV1701}
(we will call them GPPV~invariants, they are denoted by~$\hat{Z}$
and often this symbol is also used as their name). The~GPPV~invariant
is a~series in $q$ with integer coefficients and the~WRT invariant
can be recovered from the~$q\rightarrow e^{\frac{2\pi i}{k}}$ limit
(see \cite{GPV1602,GMP1605,GPPV1701,CCFGH1809,Kuch1906,Chung1906}
for details and generalisations). Physical origins of GPPV~invariants
lie in the 3d-3d~correspondence \cite{DGG1108,CCV1110,DGG1112}:
$\hat{Z}$~is a~supersymmetrix index of 3d $\mathcal{N}=2$ theory
with 2d $\mathcal{N}=(0,2)$ boundary condition studied first in~\cite{GGP1302}.
Detailed analysis of this interpretation, as well as the~application
of resurgence, can be found in~\cite{GPV1602,GMP1605,GPPV1701},
whereas \cite{CGPS1911,Chung1912} contain some recent results on
GPPV~invariants coming from the study of 3d-3d~corresponcendce.
Another interpretation of GPPV~invariants as characters of 2d~logarithmic
vertex operator algebras was proposed in \cite{CCFGH1809}. On the~other
hand, this work -- together with \cite{BMM1810,BMM1906,CFS1912}
-- initiated an~exploration of the~intruguing modular properties
of $\hat{Z}$.

GM~invariants can be treated as knot complement versions of GPPV~invariants,
$F_{K}=\hat{Z}\left(S^{3}\backslash K\right)$~\cite{GM1904} (because
of that, sometimes GM~invariants are denoted and called~$\hat{Z}$,
descending it from GPPV~invariants). From the~physical point of
view, the~GM~invariant arises from the~reduction of 6d $\mathcal{N}=(0,2)$
theory describing M5-branes on the 3-manifold with the~topology of
the~knot complement. Important properties analysed in \cite{GM1904}
include the~behaviour under surgeries, the~agreement of the~asymptotic
expansion of $F_{K}$ with the~Melvin-Morton-Rozansky expansion of
the~coloured Jones polynomials \cite{MM95,BNG96,Roz96,Roz98}, relations
to the~Alexander polynomials, and the~annihilation by the~quantum
$A$-polynomials introduced in \cite{Gar0306,Guk0306}. These relations
suggest an~interesting geometric interpretation of GM~invariants
in terms of the~annuli counting presented in \cite{GGKPS20xx} and
based on \cite{DE20xx}. Another recent developments include a~generalisation
to arbitrary gauge group \cite{Park1909}, an~adaptation of the~large
colour $R$-matrix approach to GM~invariants \cite{Park2004}, and
a~relation to Akutsu-Deguchi-Ohtsuki invariant for $q=e^{\frac{2\pi i}{k}}$
\cite{GHNPPS2005}.

From our perspective, the most important new results are closed form
expressions for $a$-deformed GM~invariants provided in \cite{GGKPS20xx}.
They reduce to the~initial $F_{K}$ for $a=q^{2}$ and turn out to
be closely related to HOMFLY-PT polynomials. This allows us to propose
a~correspondence between knot complements and quivers via GM~invariants
and quiver motivic generating series. Moreover, since HOMFLY-PT polynomials
admit $t$-deformation in terms of superpolynomials \cite{DGR0505},
we can follow \cite{GGKPS20xx} and conjecture results for $a,t$-deformed
GM~invariants. We also study the~physical intepretation of the~new
correspondence (for clarity we will refer to the~knots-quivers correspondence
of \cite{KRSS1707short,KRSS1707long} as the~standard one), generalising
the~results of \cite{EKL1811,EKL1910}. It includes an~exploration
of the~duality between 3d $\mathcal{N}=2$ theories associated to
the~knot complement and the~quiver, as well as a~study of their
BPS states, which allows us to introduce knot complement analogs of
LMOV invariants via DT invariants.

The rest of the paper is organised as follows. Section \ref{sec:Prerequisites}
provides a~short introduction to the~knots-quivers correspondence
and GM~invariants of knot complements. In section \ref{sec:Quivers-for-GM}
we state our main conjecture, assigning quivers to knot complements,
and study consequences for 3d $\mathcal{N}=2$ theories and their
BPS spectra. In section \ref{sec:Computing-the-quivers} we show how
these ideas can be applied and checked on concrete examples of unknot,
trefoil, cinquefoil, and general $T^{(2,2p+1)}$ torus knots. Section
\ref{sec:t-deformation} contains a~$t$-deformation of previous
results inspired by the~categorification of HOMFLY-PT polynomials.
In section \ref{sec:Future-directions} we conclude with a~discussion
of interesting open problems.

\section{Prerequisites\label{sec:Prerequisites}}

In this section we recall relevant aspects of the~knots-quivers correspondence,
its physical interpretation, and GM~invariants of knot complements.

\subsection{Knot invariants}

If $K\subset S^{3}$ is a~knot, then its HOMFLY-PT polynomial $P_{K}(a,q)$
\cite{HOMFLY,PT} is a topological invariant which can be calculated
via the~skein relation. More generally, coloured HOMFLY-PT polynomials
$P_{K,R}(a,q)$ are similar polynomial knot invariants depending also
on a~representation $R$ of the~Lie algebra~$\mathfrak{u}(N)$.
In this setting, the original HOMFLY-PT corresponds to the~fundamental
representation. From the~physical point of view, $P_{K,R}(a,q)$
is the~expectation value of the~knot viewed as a~Wilson line in
U$(N)$ Chern-Simons gauge theory \cite{Witten_Jones}.

In the context of the~knots-quivers correspondence, we are interested
in the HOMFLY-PT generating series:
\begin{equation}
P_{K}(\lambda,a,q)=\sum_{r=0}^{\infty}P_{K,r}(a,q)\lambda^{-r},
\end{equation}
where $P_{K,r}(a,q)$ are HOMFLY-PT polynomials coloured by the~totally
symmetric representations $S^{r}$ (with $r$ boxes in one row of
the Young diagram), which for brevity we will call simply HOMFLY-PT
polynomials. The unusual expansion variable with the~negative power
comes from the~necessity of resolving the~clash of four different
conventions present in the~literature (and avoiding the~confusion
with the~quiver variables):
\noindent \begin{center}
\begin{tabular}{lll}
\noalign{\vskip\doublerulesep}
$\bullet$ KRSS convention from \cite{KRSS1707short,KRSS1707long,GKS1504,KS1608}, &  & $\bullet$ EKL convention from \cite{EKL1811,EKL1910},\tabularnewline[\doublerulesep]
\noalign{\vskip\doublerulesep}
\noalign{\vskip\doublerulesep}
$\bullet$ FGS convention from \cite{AFGS1203,FGS1205,FGSS1209,GM1904,Park1909}, &  & $\bullet$ EGGKPS convention from \cite{GGKPS20xx}.\tabularnewline[\doublerulesep]
\noalign{\vskip\doublerulesep}
\end{tabular}
\par\end{center}

\noindent The dictionary is given by
\begin{align}
\lambda & =x_{\textrm{KRSS}}^{-1}=x_{\textrm{EKL}}^{-1}=y_{\textrm{FGS}}=y_{\textrm{EGGKPS}}, & \mu & =y_{\textrm{KRSS}}=y_{\textrm{EKL}}^{1/2}=x_{\textrm{FGS}}q_{\textrm{FGS}}^{-1}=x_{\textrm{EGGKPS}},\label{eq:different conventions}\\
a & =a_{\textrm{KRSS}}^{2}=a_{\textrm{EKL}}^{2}=a_{\textrm{FGS}}=a_{\textrm{EGGKPS}}, & q & =q_{\textrm{KRSS}}^{2}=q_{\textrm{EKL}}^{2}=q_{\textrm{FGS}}=q_{\textrm{EGGKPS}}.\nonumber 
\end{align}

HOMFLY-PT polynomials satisfy recurrence relations encoded in the~quantum
$a$-deformed $A$-polynomials \cite{AV1204,GLL1604,AFGS1203,FGS1205,FGSS1209}:
\begin{equation}
\hat{A}(\hat{\mu},\hat{\lambda},a,q)P_{K,r}(a,q)=0,
\end{equation}
where 
\begin{equation}
\hat{\mu}P_{K,r}(a,q)=q^{r}P_{K,r}(a,q),\qquad\hat{\lambda}P_{K,r}(a,q)=P_{K,r+1}(a,q).
\end{equation}
For simplicity, in the remaining part of the paper we will drop ``$a$-deformed''
and call $\hat{A}(\hat{\mu},\hat{\lambda},a,q)$ the~quantum $A$-polynomials.

The LMOV invariants \cite{OV9912,LM01,LMV00} are numbers assembled
into the~LMOV generating function $N(\lambda,a,q)=\sum_{r,i,j}N_{r,i,j}\lambda^{-r}a^{i}q^{j}$
that gives the following expression for the~HOMFLY-PT generating
series:
\begin{equation}
P_{K}(\lambda,a,q)=\mathrm{Exp}\left[\frac{N(\lambda,a,q)}{1-q}\right].\label{eq:P_K=00003DExp}
\end{equation}
Exp is the~plethystic exponential: if $\ensuremath{f(t)=\sum_{n}a_{n}t^{n}}$
and $a_{0}=0$, then
\begin{equation}
\mathrm{Exp}\left[f(t)\right]=\exp\left[\sum_{k}\tfrac{1}{k}f(t^{k})\right]=\prod_{n}(1-t^{n})^{a_{n}}.
\end{equation}
LMOV invariants can be extracted also from the $A$-polynomials, see
\cite{GKS1504,KS1608}.

\subsection{Quivers and their representations}

A~quiver $Q$ is an\,oriented graph, i.e.~a~pair $(Q_{0},Q_{1})$
where $Q_{0}$ is a~finite set of vertices and $Q_{1}$~is a~finite
set of arrows between them. We number the~vertices by $1,2,...,m=|Q_{0}|$.
An~adjacency matrix of~$Q$ is the~$m\times m$ integer matrix
with entries $C_{ij}$ equal to the~number of arrows from $i$ to~$j$.
If $C_{ij}=C_{ji}$, we call the quiver symmetric.

A~quiver representation with dimension vector $\boldsymbol{d}=(d_{1},...,d_{m})$
is the~assignment of a~vector space of dimension $d_{i}$ to the~node
$i\in Q_{0}$ and of a~linear map $\gamma_{ij}:\mathbb{C}^{d_{i}}\rightarrow\mathbb{C}^{d_{j}}$
to each arrow from vertex $i$ to vertex $j$. Quiver representation
theory studies moduli spaces of stable quiver representations. While
explicit expressions for invariants describing those spaces are hard
to find in general, they are quite well understood in the~case of
symmetric quivers \cite{KS0811,KS1006,MR1411,FR1512,Efi12}. Important
information about the~moduli space of representations of a~symmetric
quiver is encoded in the~motivic generating series defined as
\begin{equation}
P_{Q}(\boldsymbol{x},q)=\sum_{d_{1},\ldots,d_{m}\geq0}(-q^{1/2})^{\sum_{i,j}C_{ij}d_{i}d_{j}}\prod_{i=1}^{m}\frac{x_{i}^{d_{i}}}{(q;q)_{d_{i}}},
\end{equation}
where the\,denominator is the~$q$-Pochhammer symbol:
\begin{equation}
(z;q)_{n}=\prod_{k=0}^{n-1}(1-zq^{k}).
\end{equation}
If we write
\begin{equation}
P_{Q}(\boldsymbol{x},q)=\textrm{Exp}\left[\frac{\Omega(\boldsymbol{x},q)}{1-q}\right],\label{eq:P_Q=00003DExp}
\end{equation}
we obtain the~generating series of motivic Donaldson-Thomas (DT)
invariants $\Omega_{\boldsymbol{d},s}$ \cite{KS0811,KS1006}:
\begin{equation}
\Omega(\boldsymbol{x},q)=\sum_{\boldsymbol{d},s}\Omega_{\boldsymbol{d},s}\boldsymbol{x}^{\boldsymbol{d}}q^{s}=\sum_{\boldsymbol{d},s}\Omega_{(d_{1},...,d_{m}),s}\left(\prod_{i}x_{i}^{d_{i}}\right)q^{s}.
\end{equation}
The DT invariants have two geometric interpretations, either as the~intersection
homology Betti numbers of the~moduli space of all semi-simple representations
of~$Q$ of dimension vector~$\boldsymbol{d}$, or as the~Chow-Betti
numbers of the~moduli space of all simple representations of~$Q$
of dimension vector~$\boldsymbol{d}$, see~\cite{MR1411,FR1512}.
\cite{Efi12} provides a~proof of integrality of DT invariants for
the~symmetric quivers.

\subsection{Knots-quivers correspondence\label{subsec:Knots-quivers-correspondence}}

The~knots-quivers correspondence\,\cite{KRSS1707short,KRSS1707long}
is a~conjecture that for each knot~$K$ there exist a~quiver~$Q$
and integers $n_{i}$, $a_{i}$, $l_{i}$, $\ensuremath{i\in Q_{0}}$,
such that
\begin{equation}
P_{K}(\lambda,a,q)=\left.P_{Q}(\boldsymbol{x},q)\right|_{x_{i}=\lambda^{n_{i}}a^{a_{i}}q^{l_{i}}}.
\end{equation}
If we substitute (\ref{eq:P_K=00003DExp}) and (\ref{eq:P_Q=00003DExp}),
we obtain the\,knots-quivers correspondence at the~level of LMOV
and DT invariants:
\begin{equation}
N(\lambda,a,q)=\left.\Omega(\boldsymbol{x},q)\right|_{x_{i}=\lambda^{n_{i}}a^{a_{i}}q^{l_{i}}}.
\end{equation}
Since DT invariants are integer, this equation implies integrality
of $N_{r,i,j}$, which is known as the~LMOV conjecture.

From the physical point of view, DT and LMOV invariants count BPS
states in 3d $\mathcal{N}=2$ theories denoted by $T[L_{K}]$ and
$T[Q_{L_{K}}]$ \cite{EKL1811,EKL1910}. $T[L_{K}]$ is the~effective
3d $\mathcal{N}=2$ theory on the~world-volume of M5-brane wrapped
on the~knot conormal inside the~resolved conifold:
\[
\begin{split}\text{space-time}:\quad & \mathbb{R}^{4}\times S^{1}\times X\\
 & \cup\phantom{\ \times S^{1}\times\ \ }\cup\\
\text{\text{M5-brane}}:\quad & \mathbb{R}^{2}\times S^{1}\times L_{K}.
\end{split}
\]
The structure of $T[L_{K}]$ can be read from the semiclassical limit
($q=e^{\hbar}\rightarrow1$) of the HOMFLY-PT generating series:
\begin{equation}
\sum_{r=0}^{\infty}P_{K,r}(a,q)\lambda^{-r}\stackrel[\hbar\rightarrow0]{q^{r}=\mu}{\longrightarrow}\int\frac{d\mu}{\mu}\prod_{i}\frac{dz_{i}}{z_{i}}\exp\left[\frac{1}{\hbar}\widetilde{\mathcal{W}}_{T[L_{K}]}(\mu,\lambda,a,z_{i})+\mathcal{O}(\hbar^{0})\right],\label{eq:Semiclassical limit HOMFLY-PT}
\end{equation}
where the integral $\int\frac{d\mu}{\mu}\prod_{i}\frac{dz_{i}}{z_{i}}$
corresponds to the gauge group $\textrm{U}(1)_{M}\times\textrm{U}(1)_{Z_{1}}\times\ldots\times\textrm{U}(1)_{Z_{k}}$
(we single out $\textrm{U}(1)_{M}$ and its fugacity $\mu$ because
for the knot complement theory it becomes a~global symmetry). $\widetilde{\mathcal{W}}_{T[L_{K}]}(\mu,\lambda,a,z_{i})$
is the twisted superpotential with two typical kinds of contributions:

\begin{equation}
\begin{split}\textrm{Li}_{2}\ensuremath{\left(a^{n_{Q}}\mu^{n_{M}}z_{i}^{n_{Z_{i}}}\right)}\qquad & \longleftrightarrow\qquad\text{(chiral field)}\,,\\
\frac{\kappa_{ij}}{2}\log\zeta_{i}\cdot\log\zeta_{j}\qquad & \longleftrightarrow\qquad\text{(Chern-Simons coupling)}\,.
\end{split}
\label{eq:Li2 and logs dictionary}
\end{equation}
Each dilogarithm is interpreted as the~one-loop contribution of a~chiral
field with charges $(n_{Q},n_{M},n_{Z_{i}})$ under the~global symmetry
U$(1)_{Q}$ (arising from the~internal 2-cycle in the~resolved conifold
geometry) and the gauge group $\textrm{U}(1)_{M}\times\textrm{U}(1)_{Z_{1}}\times\ldots\times\textrm{U}(1)_{Z_{k}}$.
Quadratic-logarithmic terms are identified with Chern-Simons couplings
among the~various U(1) gauge and global symmetries, with $\zeta_{i}$
denoting the~respective fugacities. For more details see \cite{EKL1811,FGS1205,FGSS1209,DGH1006,TY1103,DGG1108,CCV1110}.

Integrating over the~gauge fugacities $z_{i}$ using a~saddle-point
approximation gives the~effective twisted superpotential of the~theory:
\begin{equation}
\widetilde{\mathcal{W}}_{T[L_{K}]}^{\textrm{eff}}(\mu,\lambda,a)=\widetilde{\mathcal{W}}_{T[L_{K}]}(\mu,\lambda,a,z_{i}^{*}),\qquad\text{where}\qquad\left.\frac{\partial\widetilde{\mathcal{W}}_{T[L_{K}]}(\mu,\lambda,a,z_{i})}{\partial z_{i}}\right|_{z_{i}=z_{i}^{*}}=0.
\end{equation}
The moduli space of vacua of $T[L_{K}]$, given by the~extremal points
of the~effective twisted superpotential, coincides with the~graph
of the~classical $A$-polynomial:
\begin{equation}
\begin{split}\frac{\partial\widetilde{\mathcal{W}}_{T[L_{K}]}^{\textrm{eff}}(\mu,\lambda,a)}{\partial\log\lambda}=0\qquad\Leftrightarrow\qquad A(\mu,\lambda,a) & =0,\\
A(\mu,\lambda,a) & =\underset{q\rightarrow1}{\lim}\hat{A}(\hat{\mu},\hat{\lambda},a,q).
\end{split}
\end{equation}

In analogy to $T[L_{K}]$, the~structure of $T[Q_{L_{K}}]$ is encoded
in the~semiclassical limit of the~motivic generating series \cite{EKL1811}:
\begin{equation}
\begin{split} & P_{Q}(\boldsymbol{x},q)\stackrel[\hbar\rightarrow0]{q^{d_{i}}=y_{i}}{\longrightarrow}\int\prod_{i}\frac{dy_{i}}{y_{i}}\exp\ensuremath{\left[\frac{1}{\hbar}\widetilde{\mathcal{W}}_{T[Q_{L_{K}}]}(\boldsymbol{x},\boldsymbol{y})+\mathcal{O}(\hbar^{0})\right]},\\
 & \widetilde{\mathcal{W}}_{T[Q_{L_{K}}]}(\boldsymbol{x},\boldsymbol{y})=\sum_{i}\textrm{Li}_{2}(y_{i})+\log\left(\ensuremath{(-1)^{C_{ii}}x_{i}}\right)\,\log y_{i}+\sum_{i,j}\frac{C_{ij}}{2}\log y_{i}\,\log y_{j}.
\end{split}
\label{eq:Semiclassical limit P_Q}
\end{equation}
Using the~dictionary (\ref{eq:Semiclassical limit HOMFLY-PT}-\ref{eq:Li2 and logs dictionary}),
we can interpret the elements of (\ref{eq:Semiclassical limit P_Q})
in the following way:
\begin{itemize}
\item The integral $\int\prod_{i}\frac{dy_{i}}{y_{i}}$ corresponds to having
the gauge group $\textrm{U}(1)^{(1)}\times\dots\times\textrm{U}(1)^{(m)}$,
\item $\textrm{Li}_{2}(y_{i})$ represents the~chiral field with charge
$1$ under $\textrm{U}(1)^{(i)}$,
\item $\frac{C_{ij}}{2}\log y_{i}\,\log y_{j}$ corresponds to the~gauge
Chern-Simons couplings, $\kappa_{ij}^{\textrm{eff}}=C_{ij}$,
\item $\log\left(\ensuremath{(-1)^{C_{ii}}x_{i}}\right)\,\log y_{i}$ represents
the~Chern-Simons coupling between a gauge symmetry and its dual topological
symmetry (the Fayet-Iliopoulos coupling).
\end{itemize}
The saddle point of the twisted superpotential encodes the moduli
space of vacua of $T[Q_{L_{K}}]$ and defines the quiver $A$-polynomials
\cite{EKL1811,EKL1910,PSS1802,PS1811}:
\begin{equation}
\frac{\partial\widetilde{\mathcal{W}}_{T[Q_{L_{K}}]}(\boldsymbol{x},\boldsymbol{y})}{\partial\log y_{i}}=0\qquad\Leftrightarrow\qquad A_{i}(\boldsymbol{x},\boldsymbol{y})=1-y_{i}-x_{i}(-y_{i})^{C_{ii}}\prod_{j\neq i}y_{j}^{C_{ij}}=0.
\end{equation}
$A_{i}(\boldsymbol{x},\boldsymbol{y})$ is a classical limit of the
quantum quiver $A$-polynomial, which annihilates the motivic generating
series:
\begin{equation}
\begin{split}\hat{A}_{i}(\hat{\boldsymbol{x}},\hat{\boldsymbol{y}},q)P_{Q}(\boldsymbol{x},q) & =0,\\
\hat{x}_{i}P_{Q}(x_{1},\ldots,x_{i},\ldots,x_{m},q) & =x_{i}P_{Q}(x_{1},\ldots,x_{i},\ldots,x_{m},q),\\
\hat{y}_{i}P_{Q}(x_{1},\ldots,x_{i},\ldots,x_{m},q) & =P_{Q}(x_{1},\ldots,qx_{i},\ldots,x_{m},q).
\end{split}
\end{equation}
The general formula for the quantum quiver $A$-polynomial corresponding
to the quiver with adjacency matrix $C$ is given by \cite{EKL1910}
\begin{equation}
\hat{A}_{i}(\hat{\boldsymbol{x}},\hat{\boldsymbol{y}},q)=1-\hat{y}_{i}-\hat{x}_{i}(-q^{1/2}\hat{y}_{i})^{C_{ii}}\prod_{j\neq i}\hat{y}_{j}^{C_{ij}}
\end{equation}
and we can see that
\begin{equation}
A_{i}(\boldsymbol{x},\boldsymbol{y})=\underset{q\rightarrow1}{\lim}\hat{A}_{i}(\hat{\boldsymbol{x}},\hat{\boldsymbol{y}},q).
\end{equation}

\subsection{GM invariants}

GM~invariants $F_{K}=\hat{Z}\left(S^{3}\backslash K\right)$ implicitly
depend on the~gauge group and results of \cite{GM1904} correspond
to the~simplest nontrivial case of~SU(2). General case is analysed
in \cite{Park1909}, but it is often very involved from computational
point of view. One of the~goals of \cite{GGKPS20xx} was overcoming
these difficulties and studying the~large-$N$ behaviour of $F_{K}^{\textrm{SU}(N)}(\mu,q)$
-- the~GM~invariant corresponding to the~symmetric representations
of SU($N$). It turns out that we can introduce $a$-deformed GM~invariants
$F_{K}(\mu,a,q)$ which capture all $N$ by the~following relation:
\begin{equation}
F_{K}(\mu,a=q^{N},q)=F_{K}^{\textrm{SU}(N)}(\mu,q).
\end{equation}
$F_{K}(\mu,a,q)$ is annihilated by the~quantum $a$-deformed $A$-polynomials:
\begin{equation}
\hat{A}(\hat{\mu},\hat{\lambda},a,q)F_{K}(\mu,a,q)=0,\label{eq:AF=00003D0}
\end{equation}
where 
\begin{equation}
\hat{\mu}F_{K}(\mu,a,q)=\mu F_{K}(\mu,a,q),\qquad\hat{\lambda}F_{K}(\mu,a,q)=F_{K}(q\mu,a,q).
\end{equation}
Since we adapt the convention that the~series expansion of $F_{K}$
starts from~$1$, as in~\cite{GGKPS20xx}, sometimes we have to
rescale $\hat{\lambda}$ in order to compare to the~literature.

From our point of view, the~crucial result of~\cite{GGKPS20xx}
is the connection between $a$-deformed GM~invariants and HOMFLY-PT
polynomials. In general, we expect it to be some version of the~Fourier
transform, but it is difficult to define it properly. However, in
some cases it reduces to a~simple substitution:
\begin{equation}
F_{K}(\mu,a,q)=\left.P_{K,r}(a,q)\right|_{q^{r}=\mu}.\label{eq:connection between F_K and HOMFLY-PT}
\end{equation}
It is not yet known what conditions are sufficient and which are necessary
for this equation to hold. \cite{GGKPS20xx}~contains an~explicit
check for the~trefoil by comparing with $F_{3_{1}}^{\textrm{SU}(N)}(\mu,q)$
from \cite{GM1904,Park1909}. Unfortunately, already for the~figure-eight
knot the~substitution $q^{r}=\mu$ leads to ill-defined series in
both $\mu$ and~$\mu^{-1}$ and this situtation is ubiquitous. Nevertheless,
solving equation (\ref{eq:AF=00003D0}) order by order in~$\mu$
works well for~$4_{1}$, which suggests that there exists a~well-defined
$F_{4_{1}}(\mu,a,q)$, however not equal to~$\left.P_{4_{1},r}(a,q)\right|_{q^{r}=\mu}$.
On the~other hand, there is a~class of knots for which we know general
expressions for $P_{K,r}(a,q)$ and the~substitution $q^{r}=\mu$
leads to well-defined series. They are $T^{(2,2p+1)}$~torus knots
and for them (\ref{eq:connection between F_K and HOMFLY-PT}) is conjectured
to hold \cite{GGKPS20xx}. Closed form expressions for~$F_{K}$ are
essential for finding quivers, so $T^{(2,2p+1)}$~torus knots will
be main focus of our interest.

Since there exists a~$t$-deformation of HOMFLY-PT polynomials provided
by the~superpolynomials~\cite{DGR0505}, it is natural to consider
the~following $t$-deformation of GM~invariants~\cite{GGKPS20xx}:
\begin{equation}
F_{K}(\mu,a,q,t)=\left.\mathcal{P}_{K,r}(a,q,t)\right|_{q^{r}=\mu}.\label{eq:connection between F_K and superpolynomial}
\end{equation}
We will study this deformation for $T^{(2,2p+1)}$~torus knots in
section \ref{sec:t-deformation}.

For simplicity, in the~remaining part of the paper we will refer
to $F_{K}^{\textrm{SU}(2)}(\mu,q)$, $F_{K}^{\textrm{SU}(N)}(\mu,q)$,
$F_{K}(\mu,a,q)$, and $F_{K}(\mu,a,q,t)$ broadly as GM~invariants,
specyfying the~case explicitly only when it is not obvious from the~context.

\section{Quivers for GM invariants\label{sec:Quivers-for-GM}}

In this section we introduce quivers for GM invariants and study their
physical interpretation in terms of BPS state counts and 3d $\mathcal{N}=2$
effective theories.

\subsection{Main conjecture}

The connection between GM~invariants and HOMFLY-PT polynomials, together
with closed form expressions for $F_{T^{(2,2p+1)}}(\mu,a,q)$ found
in \cite{GGKPS20xx}, suggest that the idea of knots-quivers correspondence
can be reformulated for GM~invariants. In other words, we expect
the following:

\begin{conj}

For a given knot complement $M_{K}=S^{3}\backslash K$, the~GM~invariant
$F_{K}(\mu,a,q)$ can be written in the\,form
\begin{equation}
F_{K}(\mu,a,q)=\sum_{d_{1},...,d_{m}\geq0}(-q^{1/2})^{\sum_{i,j=1}^{m}C_{ij}d_{i}d_{j}}\prod_{i=1}^{m}\frac{\mu^{n_{i}d_{i}}a^{a_{i}d_{i}}q^{l_{i}d_{i}}}{(q;q)_{d_{i}}},
\end{equation}
where $C$ is a~symmetric $m\times m$ matrix and $n_{i}$, $a_{i}$,
$l_{i}$ are fixed integers. In consequence, there exist a~quiver
$Q$ which adjacency matrix is equal to~$C$ and the~motivic generating
series 
\begin{equation}
F_{Q}(\boldsymbol{x},q)=\sum_{d_{1},\ldots,d_{m}\geq0}(-q^{1/2})^{\sum_{i,j=1}^{m}C_{ij}d_{i}d_{j}}\prod_{i=1}^{m}\frac{x_{i}^{d_{i}}}{(q;q)_{d_{i}}}
\end{equation}
reduces to the~GM invariant after the change of variables $x_{i}=\mu^{n_{i}}a^{a_{i}}q^{l_{i}}$:
\begin{equation}
F_{K}(\mu,a,q)=\left.F_{Q}(\boldsymbol{x},q)\right|_{x_{i}=\mu^{n_{i}}a^{a_{i}}q^{l_{i}}}.\label{eq:GM-Q correspondence}
\end{equation}

\end{conj}

\noindent Let us stress that we may encounter negative entries of
the quiver adjacency matrix: $C_{ij}<0$, so we can either accept
having two types of arrows (ordinary arrows and ``antiarrows'' which
annihilate each other) or use the change of framing to shift all entries
(for details see \cite{KRSS1707long}). Moreover, there is a~possiblity
of the sign difference in the knot complement and quiver variables
which would lead to more complicated change of variables $x_{i}=(-1)^{j_{i}}\mu^{n_{i}}a^{a_{i}}q^{l_{i}}$,
but for all analysed examples it was not the case.

We can also obtain quivers for SU($N$) (and more specifically initial
SU(2)) GM invariants by substituting $a=q^{N}$ in (\ref{eq:GM-Q correspondence}),
which leads to 
\begin{equation}
F_{K}^{\textrm{SU}(N)}(\mu,q)=\left.F_{Q}(\boldsymbol{x},q)\right|_{x_{i}=\mu^{n_{i}}q^{l_{i}+Na_{i}}}.
\end{equation}

\subsection{Corollary -- BPS states and 3d $\mathcal{N}=2$ effective theories\label{sec:Counting-BPS-states}}

If there exist a~quiver $Q$ corresponding to the~GM~invariant
of~$M_{K}$, we can compute the~DT~invariants using
\begin{equation}
F_{Q}(\boldsymbol{x},q)=\textrm{Exp}\left[\frac{\Omega(\boldsymbol{x},q)}{1-q}\right].\label{eq:DT invariants definition}
\end{equation}
Since $Q$ is symmetric, we immediately know that DT~invariants (which
are coefficients of $\Omega(\boldsymbol{x},q)$) are integer numbers
\cite{Efi12}.

We can also make a step back and apply the change of variables $x_{i}=\mu^{n_{i}}a^{a_{i}}q^{l_{i}}$
to (\ref{eq:DT invariants definition}). This leads us to the definition
of the knot complement analogs of LMOV invariants:
\begin{equation}
N(\mu,a,q)=\sum_{r,i,j}N_{r,i,j}\mu^{r}a^{i}q^{j}=\left.\Omega(\boldsymbol{x},q)\right|_{x_{i}=\mu^{n_{i}}a^{a_{i}}q^{l_{i}}},
\end{equation}
which implies
\begin{equation}
\begin{split}F_{K}(\mu,a,q) & =\textrm{Exp}\left[\frac{N(\mu,a,q)}{1-q}\right]=\exp\left[\sum_{n,r,i,j}\frac{N_{r,i,j}\mu^{rn}a^{in}q^{jn}}{1-q^{n}}\right]\\
 & =\prod_{r,i,j,l}\left(1-\mu^{r}a^{i}q^{j+l}\right)^{-N_{r,i,j}}=\prod_{r,i,j}(\mu^{r}a^{i}q^{j};q)_{\infty}^{-N_{r,i,j}}
\end{split}
\label{eq:F_K LMOV invariants}
\end{equation}
Since DT invariants are integer numbers, so are the~knot complement
analogs of LMOV invariants. This is consistent with the~physical
interpretation given in \cite{EKL1811,EKL1910}: we expect that $\Omega_{\boldsymbol{d},s}$
and $N_{r,i,j}$ invariants count the number of BPS states in a dual
3d $\mathcal{N}=2$ theories $T[Q_{M_{K}}]$ and $T[M_{K}]$.

We identify $T[M_{K}]$ with the~effective 3d $\mathcal{N}=2$ theory
on $\mathbb{R}^{2}\times S^{1}$ which can be engineered in two equivalent
ways. One is the~compactification of $N$~M5-branes on the~knot
complement:
\[
\begin{split}\text{space-time}:\quad & \mathbb{R}^{4}\times S^{1}\times T^{*}M_{K}\\
 & \cup\phantom{\ \times S^{1}\times\ \ }\cup\\
N~\text{M5-branes}:\quad & \mathbb{R}^{2}\times S^{1}\times M_{K}.
\end{split}
\]
When $N\rightarrow\infty$, many protected quantities -- such as
the~twisted superpotential -- depend on the~number of M5-branes
only via the~combination $q^{N}$ which can be treated as a~separate
variable $a$. The~second way comes from the large-$N$ transition
from the~deformed to the~resolved conifold \cite{GV9811,OV9912}:
\[
\begin{split}\text{space-time}:\quad & \mathbb{R}^{4}\times S^{1}\times T^{*}S^{3}\\
 & \cup\phantom{\ \times S^{1}\times\ \ }\cup\\
N~\text{M5-branes}:\quad & \mathbb{R}^{2}\times S^{1}\times S^{3}\\
\text{M5-brane}:\quad & \mathbb{R}^{2}\times S^{1}\times L_{K}
\end{split}
\longleftrightarrow\begin{split}\text{space-time}:\quad & \mathbb{R}^{4}\times S^{1}\times X\\
 & \cup\phantom{\ \times S^{1}\times\ \ }\cup\\
\text{\text{M5-brane}}:\quad & \mathbb{R}^{2}\times S^{1}\times L_{K}.
\end{split}
\]
Then $\log a$ is a~complexified K{\"a}hler parameter of $X$. More
details, together with a~description of a~third way of engineering
$T[M_{K}]$, are available in \cite{GGKPS20xx}.

The structure of $T[M_{K}]$ can be read from the semiclassical limit
of the GM invariant:
\begin{equation}
F_{K}(\mu,a,q)\underset{\hbar\rightarrow0}{\rightarrow}\int\prod_{i}\frac{dz_{i}}{z_{i}}\exp\left[\frac{1}{\hbar}\widetilde{\mathcal{W}}_{T[M_{K}]}(\mu,a,z_{i})+\mathcal{O}(\hbar^{0})\right].\label{eq:Semiclassical limit F_K}
\end{equation}
Recalling the~relation (\ref{eq:connection between F_K and HOMFLY-PT}),
we can see that $\widetilde{\mathcal{W}}_{T[M_{K}]}$ is the~same
as the~effective twisted superpotential of the~3d $\mathcal{N}=2$
theory analysed in~\cite{FGS1205,FGSS1209}. On the~other hand,
the~perspective of the~large-$N$ transition explains why $\widetilde{\mathcal{W}}_{T[M_{K}]}$
is a~Legendre transform of the~superpotential of the~theory $T[L_{K}]$
discussed in section \ref{subsec:Knots-quivers-correspondence}:

\begin{equation}
\widetilde{\mathcal{W}}_{T[L_{K}]}(\mu,\lambda,a,z_{i})=\widetilde{\mathcal{W}}_{T[M_{K}]}(\mu,a,z_{i})-\log\mu\log\lambda.
\end{equation}
We can also consider the~effective twisted superpotential
\[
\widetilde{\mathcal{W}}_{T[M_{K}]}^{\textrm{eff}}(\mu,a)=\widetilde{\mathcal{W}}_{T[M_{K}]}(\mu,a,z_{i}^{*}),\qquad\text{where}\qquad\left.\frac{\partial\widetilde{\mathcal{W}}_{T[M_{K}]}(\mu,a,z_{i})}{\partial z_{i}}\right|_{z_{i}=z_{i}^{*}}=0,
\]
and introduce $\lambda$ back as the~variable dual to $\mu$, which
is equivalent to the saddle point equation for $\widetilde{\mathcal{W}}_{T[L_{K}]}^{\textrm{eff}}$
and the~vanishing of the~$A$-polynomial:
\begin{equation}
\frac{\partial\widetilde{\mathcal{W}}_{T[M_{K}]}^{\textrm{eff}}(\mu,a)}{\partial\log\mu}=\log\lambda\qquad\Leftrightarrow\qquad\frac{\partial\widetilde{\mathcal{W}}_{T[L_{K}]}^{\textrm{eff}}(\mu,\lambda,a)}{\partial\log\lambda}=0\qquad\Leftrightarrow\qquad A(\mu,\lambda,a)=0.\label{eq:classical A polynomial}
\end{equation}
In consenquence $T[M_{K}]$ have the~same moduli space of vacua as
$T[L_{K}]$ and both are described by the~$A$-polynomial of $K$.

In analogy to (\ref{eq:Semiclassical limit P_Q}), the structure of
$T[Q_{M_{K}}]$ is encoded in the semiclassical limit of $F_{Q}(\boldsymbol{x},q)$:
\begin{equation}
\begin{split} & F_{Q}(\boldsymbol{x},q)\stackrel[\hbar\rightarrow0]{q^{d_{i}}=y_{i}}{\longrightarrow}\int\prod_{i}\frac{dy_{i}}{y_{i}}\exp\ensuremath{\left[\frac{1}{\hbar}\widetilde{\mathcal{W}}_{T[Q_{M_{K}}]}(\boldsymbol{x},\boldsymbol{y})+\mathcal{O}(\hbar^{0})\right]},\\
 & \widetilde{\mathcal{W}}_{T[Q_{M_{K}}]}(\boldsymbol{x},\boldsymbol{y})=\sum_{i}\textrm{Li}_{2}(y_{i})+\log\left(\ensuremath{(-1)^{C_{ii}}x_{i}}\right)\,\log y_{i}+\sum_{i,j}\frac{C_{ij}}{2}\log y_{i}\,\log y_{j}.
\end{split}
\label{eq:Semiclassical limit F_Q}
\end{equation}
$\widetilde{\mathcal{W}}_{T[Q_{M_{K}}]}(\boldsymbol{x},\boldsymbol{y})$
is a twisted superpotential of the theory $T[Q_{M_{K}}]$ and each
of its terms can be interpreted according to the dictionary described
in section \ref{subsec:Knots-quivers-correspondence}. The saddle
point of the twisted superpotential encodes the moduli space of vacua
of $T[Q_{M_{K}}]$ and defines the quiver $A$-polynomials:
\begin{equation}
\frac{\partial\widetilde{\mathcal{W}}_{T[Q_{M_{K}}]}(\boldsymbol{x},\boldsymbol{y})}{\partial\log y_{i}}=0\qquad\Leftrightarrow\qquad A_{i}(\boldsymbol{x},\boldsymbol{y})=0.\label{eq:GM quiver A polynomials}
\end{equation}
$A_{i}(\boldsymbol{x},\boldsymbol{y})$ is a classical limit of the
quantum quiver $A$-polynomial, which annihilates the motivic generating
series $F_{Q}(\boldsymbol{x},q)$:
\begin{equation}
A(\boldsymbol{x},\boldsymbol{y})=\underset{q\rightarrow1}{\lim}\hat{A}_{i}(\hat{\boldsymbol{x}},\hat{\boldsymbol{y}},q),\qquad\hat{A}_{i}(\hat{\boldsymbol{x}},\hat{\boldsymbol{y}},q)=1-\hat{y}_{i}-\hat{x}_{i}(-q^{1/2}\hat{y}_{i})^{C_{ii}}\prod_{j\neq i}\hat{y}_{j}^{C_{ij}}.\label{eq:quiver A polynomial classical limit and general formula}
\end{equation}
Applying the~change of variables $x_{i}=\mu^{n_{i}}a^{a_{i}}$ ($q\rightarrow1$)
together with~$\prod_{i}y_{i}=\lambda$, we can transform $A_{i}(\boldsymbol{x},\boldsymbol{y})$
into the~classical $A$-polynomial $A(\mu,\lambda,a)$ given by (\ref{eq:classical A polynomial}).

\section{Examples\label{sec:Computing-the-quivers}}

This section is devoted to the explicit results: computations of quivers,
BPS state counts, and 3d $\mathcal{N}=2$ theories for different GM
invariants of knot complements. The choice of examples on which we
check ideas from section \ref{sec:Quivers-for-GM} is determined by
the fact that the closed form formulas for $F_{K}(\mu,a,q)$ are completely
new results, so far available only for the $T^{(2,2p+1)}$ torus knot
complements \cite{GGKPS20xx}.

We start from the~unknot complement in the~unreduced normalisation.
For all other cases we use the~reduced one, which corresponds to
the~division by the unknot factor. For more details see \cite{GGKPS20xx}.\footnote{Note that what we call unreduced normalisation is called fully unreduced
in \cite{GGKPS20xx}.}

\subsection{Unknot complement}

The simplest example is the~unknot complement, for which the~GM~invariant
is given by \cite{GGKPS20xx}:
\begin{equation}
F_{0_{1}}(\mu,a,q)=\frac{(\mu q;q)_{\infty}}{(\mu a;q)_{\infty}}.\label{eq:GM invariant for unknot}
\end{equation}
Using formulas for quantum dilogarithms we can write
\begin{equation}
\begin{split}F_{0_{1}}(\mu,a,q) & =\left.F_{Q}(x_{1},x_{2},q)\right|_{x_{1}=\mu q^{1/2},\;x_{2}=\mu a},\\
F_{Q}(x_{1},x_{2},q) & =\sum_{d_{1},d_{2}\geq0}(-q^{1/2})^{d_{1}^{2}}\frac{x_{1}^{d_{1}}x_{2}^{d_{2}}}{(q;q)_{d_{1}}(q;q)_{d_{2}}}.
\end{split}
\label{eq:Motivic generating series for unknot}
\end{equation}
We can see that the~quiver for the unknot complement is given by
\begin{equation}
Q=\raisebox{-5ex}{\includegraphics[width=0.1\textwidth]{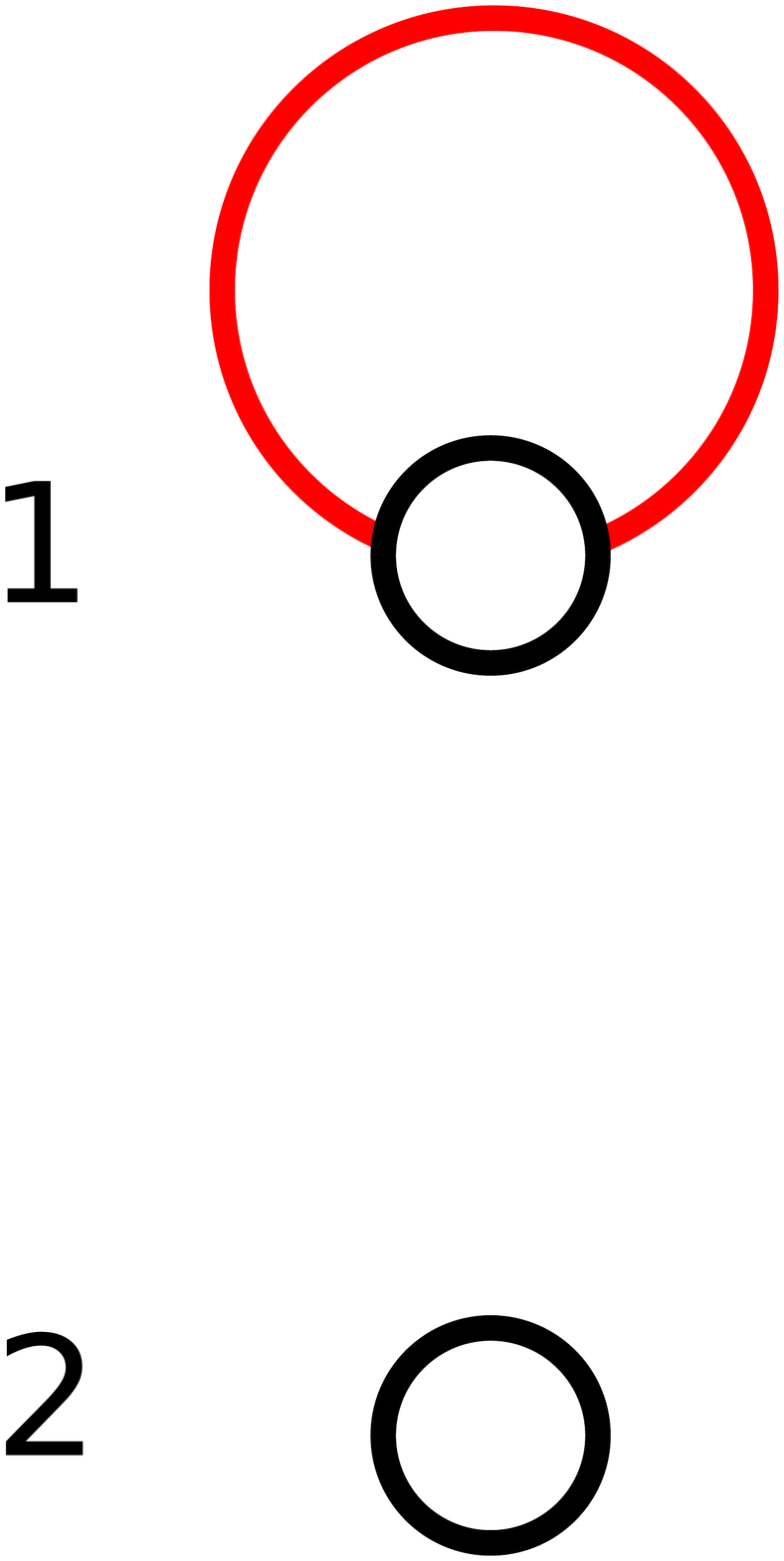}}\qquad\Longleftrightarrow\qquad C=\left[\begin{array}{cc}
1 & 0\\
0 & 0
\end{array}\right],\label{eq:Unknot quiver}
\end{equation}
which is the same as in the~standard knots-quivers correspondence
for the~unknot. However, now the~variable $a$ appears in the~change
of variables for the~node without the~loop.

For the SU($N$) GM~invariants we obtain:
\begin{equation}
F_{0_{1}}^{\textrm{SU}(N)}(\mu,q)=\frac{(\mu q;q)_{\infty}}{(\mu q^{N};q)_{\infty}}=\left.F_{Q}(x_{1},x_{2},q)\right|_{x_{1}=\mu q^{1/2},\;x_{2}=\mu q^{N}}.
\end{equation}
In case of original SU(2) GM~invariants, we still have same quiver
(\ref{eq:Unknot quiver}) and motivic generating series~(\ref{eq:Motivic generating series for unknot}),
but the~change of variables reduces to $x_{1}=\mu q^{1/2},\;x_{2}=\mu q^{2}.$

Since
\begin{equation}
F_{Q}(x_{1},x_{2},q)=\textrm{Exp}\left[\frac{\Omega(\boldsymbol{x},q)}{1-q}\right]=\textrm{Exp}\left[\frac{\sum_{\boldsymbol{d},s}\Omega_{\boldsymbol{d},s}\boldsymbol{x}^{\boldsymbol{d}}q^{s}}{1-q}\right]=\textrm{Exp}\left[\frac{-q^{1/2}x_{1}+x_{2}}{1-q}\right],
\end{equation}
we have only two nonzero DT invariants:
\begin{equation}
\Omega_{(1,0),1/2}=-1,\qquad\Omega_{(0,1),0}=1.
\end{equation}
By definition, they lead to two nonzero knot complement analogs of
LMOV invariants, exactly as in the~case of the~standard LMOV invariants
for the~unknot \cite{OV9912}:
\begin{equation}
\begin{split}N(\mu,a,q)= & \left.\Omega(\boldsymbol{x},q)\right|_{x_{1}=\mu q^{1/2},\;x_{2}=\mu a}\\
\sum_{r,i,j}N_{r,i,j}\mu^{r}a^{i}q^{j}= & -\mu q+\mu a
\end{split}
\qquad\Longrightarrow\qquad\begin{cases}
N_{1,0,1}=-1\\
N_{1,1,0}=1.
\end{cases}
\end{equation}
Alternatively, we could have obtained this result by a~direct comparison
between (\ref{eq:GM invariant for unknot}) and (\ref{eq:F_K LMOV invariants}).

As discussed in section \ref{sec:Quivers-for-GM}, we can use the~semiclassical
limit of the motivic generating series and the~GM~invariant to obtain
effective twisted superpotentials of theories $T[Q_{M_{0_{1}}}]$
and $T[M_{0_{1}}]$, whose BPS~states are counted by DT~invariants
and knot complement analogs of LMOV~invariants respectively. The~limit~(\ref{eq:Semiclassical limit F_Q})
for the~unknot complement quiver is given by
\begin{equation}
\begin{split} & F_{Q}(\boldsymbol{x},q)\stackrel[\hbar\rightarrow0]{q^{d_{i}}=y_{i}}{\longrightarrow}\int\frac{dy_{1}}{y_{1}}\frac{dy_{2}}{y_{2}}\exp\ensuremath{\left[\frac{1}{\hbar}\widetilde{\mathcal{W}}_{T[Q_{M_{0_{1}}}]}(\boldsymbol{x},\boldsymbol{y})+\mathcal{O}(\hbar^{0})\right]},\\
 & \widetilde{\mathcal{W}}_{T[Q_{M_{0_{1}}}]}(\boldsymbol{x},\boldsymbol{y})=\textrm{Li}_{2}\left(y_{1}\right)+\textrm{Li}_{2}\left(y_{2}\right)+\log(-x_{1})\log y_{1}+\log x_{2}\log y_{2}+\frac{1}{2}\log y_{1}\log y_{1}.
\end{split}
\end{equation}
We can see that $T[Q_{M_{0_{1}}}]$ is a~$\textrm{U}(1)^{(1)}\times\textrm{U}(1)^{(2)}$
gauge theory with one chiral field for each group and effective Chern-Simons
level one for $\textrm{U}(1)^{(1)}$, which is the same as $T[Q_{L_{0_{1}}}]$
\cite{EKL1811}. According to~(\ref{eq:GM quiver A polynomials}),
the~critical point of quiver twisted superpotential given by
\begin{equation}
\begin{split}0=\frac{\partial\widetilde{\mathcal{W}}_{T[Q_{M_{0_{1}}}]}}{\partial\log y_{1}} & =\log(-x_{1})+\log y_{1}-\log(1-y_{1})\,,\\
0=\frac{\partial\widetilde{\mathcal{W}}_{T[Q_{M_{0_{1}}}]}}{\partial\log y_{2}} & =\log x_{2}+\log y_{2}-\log(1-y_{2})
\end{split}
\end{equation}
defines the quiver $A$-polynomials
\begin{align}
A_{1}(\boldsymbol{x},\boldsymbol{y}) & =1-y_{1}+x_{1}y_{1}, & A_{2}(\boldsymbol{x},\boldsymbol{y}) & =1-y_{2}-x_{2}.\label{eq:classical A polynomials unknot quiver}
\end{align}
$A_{1}$ and $A_{2}$ are decoupled, which originates from the lack
of arrows between vertices $1$ and $2$ in $Q$. The~quantum quiver
$A$-polynomial, which annihilates $F_{Q}(\boldsymbol{x},q)$ and
reduces to (\ref{eq:classical A polynomials unknot quiver}) for $q\rightarrow1$,
is given by
\begin{align}
\hat{A}_{1}(\hat{\boldsymbol{x}},\hat{\boldsymbol{y}},q) & =1-\hat{y}_{1}+\hat{x}_{1}(-q^{1/2}\hat{y}_{1}), & \hat{A}_{2}(\hat{\boldsymbol{x}},\hat{\boldsymbol{y}},q) & =1-\hat{y}_{2}-\hat{x}_{2}.\label{eq:quantum A polynomials unknot quiver}
\end{align}
Applying the~change of variables $x_{1}=\mu,\;x_{2}=\mu a,\;y_{1}y_{2}=\lambda$
to (\ref{eq:classical A polynomials unknot quiver}), we recover the~classical
$A$-polynomial of the~unknot\footnote{We have to take into account the rescaling of $\lambda$ by $a^{1/2}$
with respect to $y_{\textrm{FGS}}$ due to the fact the we start $F_{0_{1}}(\mu,a,q)$
from 1.} \cite{FGS1205}:
\begin{equation}
A_{0_{1}}(\mu,\lambda,a)=1-a\mu-\lambda+\mu\lambda,\label{eq:A-polynomial unknot}
\end{equation}
which is in line with section \ref{sec:Quivers-for-GM}.

We also expect that the~$A$-polynomial of the~unknot encodes the~moduli
space of vacua of~$T[M_{0_{1}}]$ -- the~theory read from the semiclassical
limit of $F_{0_{1}}(\mu,a,q)$:
\begin{equation}
\begin{split} & F_{0_{1}}(\mu,a,q)\underset{\hbar\rightarrow0}{\rightarrow}\exp\left[\frac{1}{\hbar}\widetilde{\mathcal{W}}_{T[M_{0_{1}}]}(\mu,a)+\mathcal{O}(\hbar^{0})\right],\\
 & \widetilde{\mathcal{W}}_{T[M_{0_{1}}]}(\mu,a)=\textrm{Li}_{2}(\mu)-\textrm{Li}_{2}(a\mu).
\end{split}
\end{equation}
Solving 
\begin{equation}
\log\lambda=\frac{\partial\widetilde{\mathcal{W}}_{T[M_{0_{1}}]}(\mu,a)}{\partial\log\mu}=\log\left(1-a\mu\right)-\log\left(1-\mu\right)
\end{equation}
we indeed find an~agreement with $A_{0_{1}}(\mu,\lambda,a)=0$.

\subsection{Trefoil knot complement\label{subsec:Trefoil-knot-complement}}

Let us move to the~trefoil knot complement and start from the~closed
form expression found in \cite{GGKPS20xx}:
\begin{equation}
F_{3_{1}}(\mu,a,q)=\sum_{k_{1}=0}^{\infty}(\mu q)^{k_{1}}\frac{(\mu;q^{-1})_{k_{1}}(aq^{-1};q)_{k_{1}}}{(q;q)_{k_{1}}}.
\end{equation}
Using the~formula
\begin{equation}
(\xi;q)_{k}=\sum_{l=0}^{k}\left(-q^{1/2}\right)^{l^{2}}\xi^{l}q^{-\frac{l}{2}}\frac{(q;q)_{k}}{(q;q)_{l}(q;q)_{k-l}}
\end{equation}
for $(\mu q^{-1};q^{-1})_{k}$ and the~identity
\begin{equation}
\frac{(q^{-1};q^{-1})_{k}}{(q^{-1};q^{-1})_{l}(q^{-1};q^{-1})_{k-l}}=\left(-q^{1/2}\right)^{l^{2}+(k-l)^{2}-k^{2}}\frac{(q;q)_{k}}{(q;q)_{l}(q;q)_{k-l}},
\end{equation}
we can write
\begin{equation}
F_{3_{1}}(\mu,a,q)=\sum_{k_{1}=0}^{\infty}\sum_{l_{1}=0}^{k_{1}}\mu^{k_{1}+l_{1}}q^{k_{1}+\frac{l_{1}}{2}}\left(-q^{1/2}\right)^{(k_{1}-l_{1})^{2}-k_{1}^{2}}\frac{(aq^{-1};q)_{k_{1}}}{(q;q)_{l_{1}}(q;q)_{k_{1}-l_{1}}}.
\end{equation}
Then we use
\begin{equation}
\frac{(\xi;q)_{k}}{(q;q)_{l}(q;q)_{k-l}}=\sum_{\substack{\alpha+\beta=l\\
\gamma+\delta=k-l
}
}\xi^{\alpha+\gamma}q^{-\frac{1}{2}(\alpha+\gamma)}\left(-q^{1/2}\right)^{\alpha^{2}+\gamma^{2}+2\gamma l}
\end{equation}
to get
\begin{equation}
F_{3_{1}}(\mu,a,q)=\sum_{k_{1}=0}^{\infty}\sum_{l_{1}=0}^{k_{1}}\sum_{\substack{\alpha_{1}+\beta_{1}=l_{1}\\
\gamma_{1}+\delta_{1}=k_{1}-l_{1}
}
}\left(-q^{1/2}\right)^{(k_{1}-l_{1})^{2}-k_{1}^{2}+\alpha_{1}^{2}+\gamma_{1}^{2}+2\gamma_{1}l_{1}}\frac{\mu^{k_{1}+l_{1}}a^{\alpha_{1}+\gamma_{1}}q^{k_{1}+\frac{l_{1}}{2}-\frac{3}{2}(\alpha_{1}+\gamma_{1})}}{(q;q)_{\alpha_{1}}(q;q)_{\beta_{1}}(q;q)_{\gamma_{1}}(q;q)_{\delta_{1}}}.
\end{equation}
After the~change of variables
\begin{equation}
\begin{split}\alpha_{1}=d_{1},\qquad\beta_{1}=d_{2}, & \qquad\gamma_{1}=d_{3},\qquad\delta_{1}=d_{4}\\
l_{1}=d_{1}+d_{2}, & \qquad k_{1}=d_{1}+d_{2}+d_{3}+d_{4}
\end{split}
\label{eq:3_1 change of vars}
\end{equation}
we have
\begin{equation}
\begin{split}F_{3_{1}}(\mu,a,q) & =\left.F_{Q}(\boldsymbol{x},q)\right|_{x_{i}=\mu^{n_{i}}a^{a_{i}}q^{l_{i}}},\\
F_{Q}(\boldsymbol{x},q) & =\sum_{d_{1},...,d_{4}\geq0}(-q^{1/2})^{-d_{2}^{2}+d_{3}^{2}-2d_{1}d_{2}-2d_{1}d_{4}-2d_{2}d_{4}}\prod_{i=1}^{4}\frac{x_{i}^{d_{i}}}{(q;q)_{d_{i}}},
\end{split}
\label{eq:Motivic generating series for trefoil}
\end{equation}
with $x_{i}=\mu^{n_{i}}a^{a_{i}}q^{l_{i}}$ given by
\begin{equation}
x_{1}=\mu^{2}a,\qquad x_{2}=\mu^{2}q^{3/2},\qquad x_{3}=\mu aq^{-1/2},\qquad x_{4}=\mu q.\label{eq:3_1 quiver change of vars}
\end{equation}
The~quiver adjacency matrix reads:
\begin{equation}
C=\left[\begin{array}{cccc}
0 & -1 & 0 & -1\\
-1 & -1 & 0 & -1\\
0 & 0 & 1 & 0\\
-1 & -1 & 0 & 0
\end{array}\right].\label{eq:adjacency matrix trefoil quiver}
\end{equation}
This form will be useful for obtaining quivers for all $T^{(2,2p+1)}$
torus knots, but we can shift the~framing by $1$ to have only the~non-negative
entries:
\begin{equation}
C=\left[\begin{array}{cccc}
1 & 0 & 1 & 0\\
0 & 0 & 1 & 0\\
1 & 1 & 2 & 1\\
0 & 0 & 1 & 1
\end{array}\right]\qquad\Longleftrightarrow\qquad Q=\raisebox{-7ex}{\includegraphics[width=0.2\textwidth]{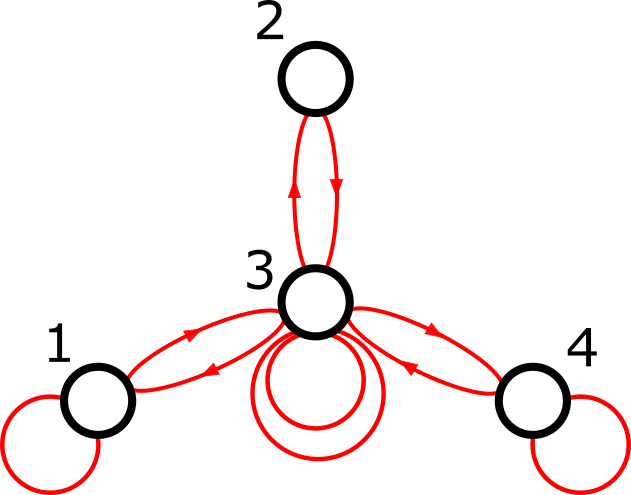}}.\label{eq:adjacency matrix trefoil quiver-non-negative}
\end{equation}

For the~trefoil the~DT spectrum is infinite, but we can find the
invariants order by order in $x_{i}$. Let us focus on the~linear
one:
\begin{equation}
\begin{split}F_{Q}(\boldsymbol{x},q) & =\textrm{Exp}\left[\frac{\Omega(\boldsymbol{x},q)}{1-q}\right]\\
\sum_{d_{1},...,d_{4}\geq0}(-q^{1/2})^{-d_{2}^{2}+d_{3}^{2}-2d_{1}d_{2}-2d_{1}d_{4}-2d_{2}d_{4}}\prod_{i=1}^{4}\frac{x_{i}^{d_{i}}}{(q;q)_{d_{i}}} & =\exp\left[\sum_{n,\boldsymbol{d},s}\frac{\Omega_{\boldsymbol{d},s}\boldsymbol{x}^{\boldsymbol{d}n}q^{sn}}{1-q^{n}}\right]\\
1+\frac{x_{1}+(-q^{1/2})^{-1}x_{2}+(-q^{1/2})x_{3}+x_{4}}{1-q}+\mathcal{O}(x_{i}^{2}) & =1+\frac{\sum_{\boldsymbol{d},s}\Omega_{\boldsymbol{d},s}\boldsymbol{x}^{\boldsymbol{d}}q^{s}}{1-q}+\mathcal{O}(x_{i}^{2}).
\end{split}
\end{equation}
In consequence
\begin{equation}
\Omega_{(1,0,0,0),0}=1,\qquad\Omega_{(0,1,0,0),-1/2}=-1,\qquad\Omega_{(0,0,1,0),1/2}=-1,\qquad\Omega_{(0,0,0,1),0}=1.
\end{equation}
Applying the~change of variables $x_{1}=\mu^{2}a,\;x_{2}=\mu^{2}q^{3/2},\;x_{3}=\mu aq^{-1/2},\;x_{4}=\mu q$,
we obtain the~(first four) knot complement analogs of LMOV~invariants
for the~trefoil knot complement:
\begin{equation}
\begin{split}N(\mu,a,q) & =\left.\Omega(\boldsymbol{x},q)\right|_{(\text{\ref{eq:3_1 quiver change of vars})}}\\
\sum_{r,i,j}N_{r,i,j}\mu^{r}a^{i}q^{j} & =\mu^{2}a-\mu^{2}q-\mu a+\mu q
\end{split}
\qquad\Longrightarrow\qquad\begin{cases}
N_{1,0,1}=1, & N_{2,0,1}=-1,\\
N_{1,1,0}=-1, & N_{2,1,0}=1.
\end{cases}
\end{equation}

$\Omega_{\boldsymbol{d},s}$ and $N_{r,i,j}$ count the BPS~states
in 3d $\mathcal{N}=2$ theories $T[Q_{M_{3_{1}}}]$ and $T[M_{3_{1}}]$
respectively. The structure of the~first theory can be read off from
the semiclassical limit of $F_{Q}(\boldsymbol{x},q)$:
\begin{align}
 & F_{Q}(\boldsymbol{x},q)\stackrel[\hbar\rightarrow0]{q^{d_{i}}=y_{i}}{\longrightarrow}\int\prod_{i=1}^{4}\frac{dy_{i}}{y_{i}}\exp\ensuremath{\left[\frac{1}{\hbar}\widetilde{\mathcal{W}}_{T[Q_{M_{3_{1}}}]}(\boldsymbol{x},\boldsymbol{y})+\mathcal{O}(\hbar^{0})\right]},\\
 & \widetilde{\mathcal{W}}_{T[Q_{M_{3_{1}}}]}(\boldsymbol{x},\boldsymbol{y})=\sum_{i=1}^{4}\textrm{Li}_{2}\left(y_{i}\right)+\log x_{1}\log y_{1}+\log(-x_{2})\log y_{2}+\log(-x_{3})\log y_{3}+\log x_{4}\log y_{4}\nonumber \\
 & \phantom{\widetilde{\mathcal{W}}_{T[Q_{M_{3_{1}}}]}(\boldsymbol{x},\boldsymbol{y})=}-\log y_{1}\log y_{2}-\frac{1}{2}\log y_{2}\log y_{2}+\frac{1}{2}\log y_{3}\log y_{3}-\log y_{1}\log y_{4}-\log y_{2}\log y_{4}.\nonumber 
\end{align}
Using the~dictionary (\ref{eq:Semiclassical limit HOMFLY-PT}-\ref{eq:Li2 and logs dictionary}),
we can see that the gauge group of $T[Q_{M_{3_{1}}}]$ is $\textrm{U}(1)^{(1)}\times\textrm{U}(1)^{(2)}\times\textrm{U}(1)^{(3)}\times\textrm{U}(1)^{(4)}$
and we have four chiral fields $\phi_{i}$ with charges $\delta_{ij}$
under $U(1)^{(j)}$. We also have gauge Chern-Simons couplings determined
by the~quiver adjacency matrix: $\kappa_{ij}^{\textrm{eff}}=C_{ij}$,
and Fayet-Iliopoulos couplings between each gauge symmetry and its
dual topological symmetry. Comparing with \cite{EKL1811}, we can
see that the~structure of $T[Q_{M_{3_{1}}}]$ is different than $T[Q_{L_{3_{1}}}]$.
The moduli space of vacua of $T[Q_{M_{3_{1}}}]$ is given by the~zero
locus of the following quiver $A$-polynomials:
\begin{align}
A_{1}(\boldsymbol{x},\boldsymbol{y}) & =1-y_{1}-x_{1}y_{2}^{-1}y_{4}^{-1}, & A_{2}(\boldsymbol{x},\boldsymbol{y}) & =1-y_{2}+x_{2}y_{1}^{-1}y_{2}^{-1}y_{4}^{-1},\\
A_{3}(\boldsymbol{x},\boldsymbol{y}) & =1-y_{3}+x_{3}y_{3}, & A_{4}(\boldsymbol{x},\boldsymbol{y}) & =1-y_{4}-x_{4}y_{1}^{-1}y_{2}^{-1},\nonumber 
\end{align}
which are the~classical limits of the~annihilators of $F_{Q}(\boldsymbol{x},q)$:
\begin{align}
\hat{A}_{1}(\hat{\boldsymbol{x}},\hat{\boldsymbol{y}},q) & =1-\hat{y}_{1}-\hat{x}_{1}\hat{y}_{2}^{-1}\hat{y}_{4}^{-1}, & \hat{A}_{2}(\hat{\boldsymbol{x}},\hat{\boldsymbol{y}},q) & =1-\hat{y}_{2}+q^{-1/2}\hat{x}_{2}\hat{y}_{1}^{-1}\hat{y}_{2}^{-1}\hat{y}_{4}^{-1},\\
\hat{A}_{3}(\hat{\boldsymbol{x}},\hat{\boldsymbol{y}},q) & =1-\hat{y}_{3}+q^{1/2}\hat{x}_{3}\hat{y}_{3}, & \hat{A}_{4}(\hat{\boldsymbol{x}},\hat{\boldsymbol{y}},q) & =1-\hat{y}_{4}-\hat{x}_{4}\hat{y}_{1}^{-1}\hat{y}_{2}^{-1}.\nonumber 
\end{align}

The structure of the~theory $T[M_{3_{1}}]$ is encoded in the~semiclassical
limit of $F_{3_{1}}(\mu,a,q)$:
\begin{equation}
\begin{split} & F_{3_{1}}(\mu,a,q)\underset{\hbar\rightarrow0}{\rightarrow}\exp\int\frac{dz}{z}\left[\frac{1}{\hbar}\widetilde{\mathcal{W}}_{T[M_{3_{1}}]}(\mu,a,z)+\mathcal{O}(\hbar^{0})\right],\\
 & \widetilde{\mathcal{W}}_{T[M_{3_{1}}]}(\mu,a,z)=\log\mu\log z-\textrm{Li}_{2}(\mu)+\textrm{Li}_{2}(\mu z^{-1})+\textrm{Li}_{2}(a)-\textrm{Li}_{2}(az)+\textrm{Li}_{2}(z).
\end{split}
\end{equation}
Extremalisation with respect to $z$ and the~introduction of the~variable
$\lambda$ dual to $\mu$ leads to
\begin{equation}
\begin{cases}
0=\frac{\partial\widetilde{\mathcal{W}}_{T[M_{3_{1}}]}(\mu,a,z)}{\partial z}\\
\log\lambda=\frac{\partial\widetilde{\mathcal{W}}_{T[M_{3_{1}}]}(\mu,a,z)}{\partial\log\mu}
\end{cases}\Longrightarrow\begin{cases}
1=\frac{\mu\left(1-\mu(z^{*})^{-1}\right)\left(1-az^{*}\right)}{\left(1-z^{*}\right)}\\
\lambda=\frac{z^{*}(1-\mu)}{1-\mu(z^{*})^{-1}}.
\end{cases}
\end{equation}
Eliminating $z^{*}$ we obtain the~zero locus of the~$A$-polynomial
\begin{equation}
A_{3_{1}}(\mu,\lambda,a)=(\mu-1)\mu^{3}-\left(1-\mu+2(1-a)\mu^{2}-a\mu^{3}+a^{2}\mu^{4}\right)\lambda+(1-a\mu)\lambda^{2},
\end{equation}
which agrees with \cite{FGS1205} after taking into account the~rescaling
of $\lambda$ by $a^{-1}$.

\subsection{Cinquefoil knot complement}

For more complicated knot complements all expressions become more
involved, so we focus on finding the~corresponding quivers, having
in mind that the~analysis of BPS~states and 3d $\mathcal{N}=2$
theories can be done analogously to the~unknot and trefoil complement
case. In this section we concentrate on the~$5_{1}=T^{(2,5)}$ knot
complement with a~plan of generalisation to all $T^{(2,2p+1)}$~torus
knot complements.

We start from the~formula for GM~invariant given in \cite{GGKPS20xx}:
\begin{equation}
F_{5_{1}}(\mu,a,q)=\sum_{0\leq k_{2}\leq k_{1}}\mu^{k_{1}+2k_{2}}q^{(k_{1}+k_{2})-k_{1}k_{2}}\frac{(aq^{-1};q)_{k_{1}}(\mu;q^{-1})_{k_{1}}}{(q;q)_{k_{1}}}\left[\begin{array}{c}
k_{1}\\
k_{2}
\end{array}\right],
\end{equation}
where we use the~$q$-binomial:
\begin{equation}
\left[\begin{array}{c}
k\\
l
\end{array}\right]=\frac{(q;q)_{k}}{(q;q)_{l}(q;q)_{k-l}}.
\end{equation}
Following the~steps from section \ref{subsec:Trefoil-knot-complement},
we obtain
\begin{align}
F_{5_{1}}(\mu,a,q)=\sum_{0\leq k_{2}\leq k_{1}}\sum_{l_{1}=0}^{k_{1}}\sum_{\substack{\alpha_{1}+\beta_{1}=l_{1}\\
\gamma_{1}+\delta_{1}=k_{1}-l_{1}
}
} & \left(-q^{1/2}\right)^{-2k_{1}k_{2}+(k_{1}-l_{1})^{2}-k_{1}^{2}+\alpha_{1}^{2}+\gamma_{1}^{2}+2\gamma_{1}l_{1}}\nonumber \\
\times & \frac{\mu^{k_{1}+2k_{2}+l_{1}}a^{\alpha_{1}+\gamma_{1}}q^{k_{1}+k_{2}+\frac{1}{2}l_{1}-\frac{3}{2}(\alpha_{1}+\gamma_{1})}}{(q;q)_{\alpha_{1}}(q;q)_{\beta_{1}}(q;q)_{\gamma_{1}}(q;q)_{\delta_{1}}}\left[\begin{array}{c}
k_{1}\\
k_{2}
\end{array}\right].
\end{align}
Now we use the~formula
\begin{equation}
\left[\begin{array}{c}
n_{1}\\
n_{2}
\end{array}\right]=\sum_{m_{2}=0}^{n_{2}}q^{(m_{1}-m_{2})(n_{2}-m_{2})}\left[\begin{array}{c}
m_{1}\\
m_{2}
\end{array}\right]\left[\begin{array}{c}
n_{1}-m_{1}\\
n_{2}-m_{2}
\end{array}\right]
\end{equation}
in two iterations:
\begin{align}
\left[\begin{array}{c}
k_{1}\\
k_{2}
\end{array}\right]=\sum_{l_{2}=0}^{k_{2}}q^{(l_{1}-l_{2})(k_{2}-l_{2})}\left[\begin{array}{c}
l_{1}\\
l_{2}
\end{array}\right]\left[\begin{array}{c}
k_{1}-l_{1}\\
k_{2}-l_{2}
\end{array}\right]=\sum_{l_{2}=0}^{k_{2}} & q^{(l_{1}-l_{2})(k_{2}-l_{2})}\sum_{\alpha_{2}+\beta_{2}=l_{2}}q^{(\alpha_{1}-\alpha_{2})(l_{2}-\alpha_{2})}\left[\begin{array}{c}
\alpha_{1}\\
\alpha_{2}
\end{array}\right]\left[\begin{array}{c}
\beta_{1}\\
\beta_{2}
\end{array}\right]\nonumber \\
\times & \sum_{\gamma_{2}+\delta_{2}=k_{2}-l_{2}}q^{(\gamma_{1}-\gamma_{2})(k_{2}-l_{2}-\gamma_{2})}\left[\begin{array}{c}
\gamma_{1}\\
\gamma_{2}
\end{array}\right]\left[\begin{array}{c}
\delta_{1}\\
\delta_{2}
\end{array}\right]\label{eq:q-binomial formula for 5_1}
\end{align}
to get
\begin{align}
F_{5_{1}}(\mu,a,q)=\sum_{0\leq k_{2}\leq k_{1}} & \sum_{l_{1}=0}^{k_{1}}\sum_{\substack{\alpha_{1}+\beta_{1}=l_{1}\\
\gamma_{1}+\delta_{1}=k_{1}-l_{1}
}
}\left(-q^{1/2}\right)^{-2k_{1}k_{2}+(k_{1}-l_{1})^{2}-k_{1}^{2}+\alpha_{1}^{2}+\gamma_{1}^{2}+2\gamma_{1}l_{1}}\\
 & \sum_{l_{2}=0}^{k_{2}}\sum_{\substack{\alpha_{2}+\beta_{2}=l_{2}\\
\gamma_{2}+\delta_{2}=k_{2}-l_{2}
}
}\left(-q^{1/2}\right)^{2(l_{1}-l_{2})(k_{2}-l_{2})+2(\alpha_{1}-\alpha_{2})(l_{2}-\alpha_{2})+2(\gamma_{1}-\gamma_{2})(k_{2}-l_{2}-\gamma_{2})}\nonumber \\
 & \times\frac{\mu^{k_{1}+2k_{2}+l_{1}}a^{\alpha_{1}+\gamma_{1}}q^{k_{1}+k_{2}+\frac{1}{2}l_{1}-\frac{3}{2}(\alpha_{1}+\gamma_{1})}}{(q;q)_{\alpha_{1}-\alpha_{2}}(q;q)_{\beta_{1}-\beta_{2}}(q;q)_{\gamma_{1}-\gamma_{2}}(q;q)_{\delta_{1}-\delta_{2}}(q;q)_{\alpha_{2}}(q;q)_{\beta_{2}}(q;q)_{\gamma_{2}}(q;q)_{\delta_{2}}}.\nonumber 
\end{align}
The change of variables
\begin{align}
\alpha_{1}= & d_{1}+d_{5}, & \beta_{1}= & d_{2}+d_{6}, & \gamma_{1}= & d_{3}+d_{7}, & \delta_{1}= & d_{4}+d_{8},\nonumber \\
\alpha_{2}= & d_{5}, & \beta_{2}= & d_{6}, & \gamma_{2}= & d_{7}, & \delta_{2}= & d_{8}\label{eq:5_1 change of vars}
\end{align}
\begin{align*}
l_{1}= & (d_{1}+d_{5})+(d_{2}+d_{6}), & k_{1}= & (d_{1}+d_{5})+(d_{2}+d_{6})+(d_{3}+d_{7})+(d_{4}+d_{8})\\
l_{2}= & d_{5}+d_{6}, & k_{2}= & d_{5}+d_{6}+d_{7}+d_{8}
\end{align*}
leads to
\begin{equation}
\begin{split}F_{5_{1}}(\mu,a,q) & =\left.F_{Q}(\boldsymbol{x},q)\right|_{x_{i}=\mu^{n_{i}}a^{a_{i}}q^{l_{i}}},\\
F_{Q}(\boldsymbol{x},q) & =\sum_{d_{1},...,d_{8}\geq0}(-q^{1/2})^{\sum_{i,j=1}^{8}C_{ij}d_{i}d_{j}}\prod_{i=1}^{8}\frac{x_{i}^{d_{i}}}{(q;q)_{d_{i}}},
\end{split}
\label{eq:Motivic generating series for cinquefoil}
\end{equation}
with $x_{i}=\mu^{n_{i}}a^{a_{i}}q^{l_{i}}$ given by
\begin{align}
x_{1}= & \mu^{2}a, & x_{2}= & \mu^{2}q^{3/2}, & x_{3}= & \mu aq^{-1/2}, & x_{4}= & \mu q,\label{eq:5_1 quiver change of vars}\\
x_{5}= & \mu^{4}aq, & x_{6}= & \mu^{4}q^{5/2}, & x_{7}= & \mu^{3}aq^{1/2}, & x_{8}= & \mu^{3}q^{2},\nonumber 
\end{align}
and
\begin{equation}
C=\left[\begin{array}{cccccccc}
0 & -1 & 0 & -1 & -1 & -1 & 0 & -1\\
-1 & -1 & 0 & -1 & -2 & -2 & 0 & -1\\
0 & 0 & 1 & 0 & -1 & -1 & 0 & 0\\
-1 & -1 & 0 & 0 & -2 & -2 & -1 & -1\\
-1 & -2 & -1 & -2 & -2 & -3 & -2 & -3\\
-1 & -2 & -1 & -2 & -3 & -3 & -2 & -3\\
0 & 0 & 0 & -1 & -2 & -2 & -1 & -2\\
-1 & -1 & 0 & -1 & -3 & -3 & -2 & -2
\end{array}\right].\label{eq:quiver for 5_1}
\end{equation}
This is the adjacency matrix corresponding to the~cinquefoil knot
complement. Performed computations and the~matrix itself may seem
invloved, but they can be considered as a~simple generalisation of
the~trefoil knot complement case. The~most important is the~fact
that the~change of variables (\ref{eq:5_1 change of vars}) can be
obtained from (\ref{eq:3_1 change of vars}). For indices $k_{1},$$l_{1},$$\alpha_{1},$$\beta_{1},$$\gamma_{1},$$\delta_{1}$,
we just substitute $d_{i}\mapsto d_{i}+d_{i+4}$. In case of $k_{2},$$l_{2},$$\alpha_{2},$$\beta_{2},$$\gamma_{2},$$\delta_{2}$,
we start from $k_{1},$$l_{1},$$\alpha_{1},$$\beta_{1},$$\gamma_{1},$$\delta_{1}$
and shift $d_{i}\mapsto d_{i+4}$. From the~point of view of the
quiver adjacency matrix, this corresponds to copying the~$4\times4$
matrix $C_{3_{1}}$ to fill $8\times8$ entries:
\begin{equation}
\left[C_{3_{1}}\right]\mapsto\left[\begin{array}{cc}
C_{3_{1}} & C_{3_{1}}\\
C_{3_{1}}^{T} & C_{3_{1}}
\end{array}\right].\label{eq:initial copying of C_3_1}
\end{equation}
We still need to include $q^{-k_{1}k_{2}}$ and the~$q$-binomial
$\left[\begin{array}{c}
k_{1}\\
k_{2}
\end{array}\right]$ which, according to (\ref{eq:q-binomial formula for 5_1}) and (\ref{eq:5_1 change of vars}),
contribute to $(-q^{1/2})^{\sum_{i,j}C_{ij}d_{i}d_{j}}$ by 
\begin{align}
\left(-q^{1/2}\right)^{2(l_{1}-l_{2})(k_{2}-l_{2})+2(\alpha_{1}-\alpha_{2})(l_{2}-\alpha_{2})+2(\gamma_{1}-\gamma_{2})(k_{2}-l_{2}-\gamma_{2})}= & \left(-q^{1/2}\right)^{2(d_{1}+d_{2})(d_{7}+d_{8})+2d_{1}d_{6}+2d_{3}d_{8}}\\
\left(-q^{1/2}\right)^{-2k_{1}k_{2}}= & \left(-q^{1/2}\right)^{-2(d_{1}+d_{2}+\ldots+d_{8})(d_{5}+d_{6}+d_{7}+d_{8})}.\nonumber 
\end{align}
In consequence we have to modify (\ref{eq:initial copying of C_3_1})
to
\begin{equation}
\left[C_{3_{1}}\right]\mapsto\left[\begin{array}{cc}
D_{0} & R_{0}\\
R_{0}^{T} & D_{1}
\end{array}\right],\label{eq:proper copying of C_3_1}
\end{equation}
where
\begin{equation}
D_{0}=C_{3_{1}},\qquad D_{1}=D_{0}-\left[\begin{array}{cccc}
2 & 2 & 2 & 2\\
2 & 2 & 2 & 2\\
2 & 2 & 2 & 2\\
2 & 2 & 2 & 2
\end{array}\right],\qquad R_{0}=D_{0}-\left[\begin{array}{cccc}
1 & 0 & 0 & 0\\
1 & 1 & 0 & 0\\
1 & 1 & 1 & 0\\
1 & 1 & 1 & 1
\end{array}\right].\label{eq:correction in copying of C_3_1}
\end{equation}
We can see that this is consistent with (\ref{eq:quiver for 5_1}).
The change of variables (\ref{eq:5_1 quiver change of vars}) is a~direct
consequence of (\ref{eq:3_1 quiver change of vars}) and (\ref{eq:5_1 change of vars}),
taking into account that in $F_{5_{1}}(\mu,a,q)$ we have an~extra
factor of $\mu^{2k_{2}}q^{k_{2}}$.

\subsection{General $T^{(2,2p+1)}$ torus knot complements\label{subsec:General--torus}}

Now we are ready to move to the~general case using recursion. We
assume that we know the~quiver for $T^{(2,2p+1)}$ torus knot complement
and look for the~quiver for $T^{(2,2(p+1)+1)}=T^{(2,2p+3)}$. The~general
formula is given by \cite{GGKPS20xx} 
\begin{equation}
\begin{split}F_{T^{(2,2p+1)}}(\mu,a,q)=\sum_{0\leq k_{p}\leq\ldots\leq k_{1}} & \mu^{2(k_{1}+\ldots+k_{p})-k_{1}}q^{(k_{1}+k_{2}+\ldots+k_{p})-\sum_{i=2}^{p}k_{i-1}k_{i}}\\
\times & \frac{(aq^{-1};q)_{k_{1}}(\mu;q^{-1})_{k_{1}}}{(q;q)_{k_{1}}}\left[\begin{array}{c}
k_{1}\\
k_{2}
\end{array}\right]\cdots\left[\begin{array}{c}
k_{p-1}\\
k_{p}
\end{array}\right],
\end{split}
\end{equation}
so when we go from $p$ to $p+1$, the~summand in $F_{T^{(2,2p+1)}}$
is multiplied by 
\begin{equation}
\mu^{2k_{p+1}}q^{k_{p+1}}(-q^{1/2})^{2k_{p}k_{p+1}}\left[\begin{array}{c}
k_{p}\\
k_{p+1}
\end{array}\right].\label{eq:induction step multiplication}
\end{equation}
This means that we copy the~rightmost (also downmost and diagonal)
$4\times4$ matrix blocks right (respectively down and further on
the diagonal), exactly like we did for $C_{3_{1}}\mapsto C_{5_{1}}$,
with the~same modification as in (\ref{eq:proper copying of C_3_1}-\ref{eq:correction in copying of C_3_1}).
Therefore the~quiver for $T^{(2,2p+1)}$ torus knot is given by
\begin{equation}
C=\left[\begin{array}{cccccc}
D_{0} & R_{0} & R_{0} & \ldots & R_{0} & R_{0}\\
R_{0}^{T} & D_{1} & R_{1} & \ldots & R_{1} & R_{1}\\
R_{0}^{T} & R_{1}^{T} & D_{2} & \ldots & R_{2} & R_{2}\\
\vdots & \vdots & \vdots & \ddots & \vdots & \vdots\\
R_{0}^{T} & R_{1}^{T} & R_{2}^{T} & \ldots & D_{p-2} & R_{p-2}\\
R_{0}^{T} & R_{1}^{T} & R_{2}^{T} & \ldots & R_{p-2}^{T} & D_{p-1}
\end{array}\right],\label{eq:torus knots quiver}
\end{equation}
where
\begin{equation}
\begin{split}D_{n} & =\left[\begin{array}{cccc}
0 & -1 & 0 & -1\\
-1 & -1 & 0 & -1\\
0 & 0 & 1 & 0\\
-1 & -1 & 0 & 0
\end{array}\right]-n\left[\begin{array}{cccc}
2 & 2 & 2 & 2\\
2 & 2 & 2 & 2\\
2 & 2 & 2 & 2\\
2 & 2 & 2 & 2
\end{array}\right]=\left[\begin{array}{cccc}
-2n & -2n-1 & -2n & -2n-1\\
-2n-1 & -2n-1 & -2n & -2n-1\\
-2n & -2n & -2n+1 & -2n\\
-2n-1 & -2n-1 & -2n & -2n
\end{array}\right],\\
R_{n} & =D_{n}-\left[\begin{array}{cccc}
1 & 0 & 0 & 0\\
1 & 1 & 0 & 0\\
1 & 1 & 1 & 0\\
1 & 1 & 1 & 1
\end{array}\right]=\left[\begin{array}{cccc}
-2n-1 & -2n-1 & -2n & -2n-1\\
-2n-2 & -2n-2 & -2n & -2n-1\\
-2n-1 & -2n-1 & -2n & -2n\\
-2n-2 & -2n-2 & -2n-1 & -2n-1
\end{array}\right].
\end{split}
\label{eq:torus knots quiver elements}
\end{equation}
We can see that, similarly to the~standard knots-quivers correspondence
\cite{KRSS1707short,KRSS1707long}, increasing $p$ does not change
any previously determined entries of the~matrix, so it makes sense
to consider the~limit $p\rightarrow\infty$ leading to an~infinite
quiver.

The change of variables $x_{i}=\mu^{n_{i}}a^{a_{i}}q^{l_{i}}$ for
$T^{(2,2p+1)}$ torus knot complement comes directly from the~generalisation
of (\ref{eq:3_1 quiver change of vars}, \ref{eq:5_1 quiver change of vars})
according to (\ref{eq:induction step multiplication}) and is given
by
\begin{equation}
\begin{split}\sum_{i}n_{i}d_{i}= & 2(d_{1}+d_{2})+1(d_{3}+d_{4})\\
+ & 4(d_{5}+d_{6})+3(d_{7}+d_{8})\\
\vdots\\
+ & 2p(d_{4p-3}+d_{4p-2})+(2p-1)(d_{4p-1}+d_{4p}),\\
\\
\sum_{i}a_{i}d_{i}= & d_{1}+d_{3}\\
+ & d_{5}+d_{7}\\
\vdots\\
+ & d_{4p-3}+d_{4p-1},
\end{split}
\label{eq:torus knots quiver change of vars}
\end{equation}
\[
\begin{split}\sum_{i}l_{i}d_{i}= & 0d_{1}+\frac{3}{2}d_{2}-\frac{1}{2}d_{3}+1d_{4}\\
+ & 1d_{5}+\frac{5}{2}d_{6}+\frac{1}{2}d_{7}+2d_{8}\\
 & \vdots\\
+ & (p-1)d_{4p-3}+\frac{2p+1}{2}d_{4p-2}+\frac{2p-3}{2}d_{4p-1}+pd_{4p}.
\end{split}
\]
We can summarise (\ref{eq:torus knots quiver}-\ref{eq:torus knots quiver change of vars})
in the concise form of the~correspondence between $T^{(2,2p+1)}$
torus knot complements and quivers:
\begin{equation}
\begin{split}F_{T^{(2,2p+1)}}(\mu,a,q)= & \left.F_{Q}(\boldsymbol{x},q)\right|_{x_{i}=\mu^{n_{i}}a^{a_{i}}q^{l_{i}}}\\
= & \left.\sum_{d_{1},...,d_{4p}\geq0}(-q^{1/2})^{\sum_{i,j=1}^{4p}C_{ij}d_{i}d_{j}}\prod_{i=1}^{4p}\frac{x_{i}^{d_{i}}}{(q;q)_{d_{i}}}\right|_{x_{i}=\mu^{n_{i}}a^{a_{i}}q^{l_{i}}.}
\end{split}
\end{equation}

\section{$\boldsymbol{t}$-deformation \label{sec:t-deformation}}

In this section we propose a~$t$-deformation of the results from
sections \ref{sec:Quivers-for-GM}-\ref{sec:Computing-the-quivers}
using $F_{K}(\mu,a,q,t)$ given in~\cite{GGKPS20xx}. These results
are based on the~$t$-deformation of GM invariants following the~generalisation
of HOMFLY-PT polynomials to superpolynomials, which is reflected in
equations (\ref{eq:connection between F_K and HOMFLY-PT}) and (\ref{eq:connection between F_K and superpolynomial}).

\subsection{Quivers}

The connection between GM~invariants and superpolynomials \cite{GGKPS20xx}
suggest the~following $t$-deformed correspondence:

\begin{conj}

For a given knot complement $M_{K}=S^{3}\backslash K$, the~GM~invariant
$F_{K}(\mu,a,q,t)$ can be written in the\,form
\begin{equation}
F_{K}(\mu,a,q,t)=\sum_{d_{1},...,d_{m}\geq0}(-q^{1/2})^{\sum_{i,j=1}^{m}C_{ij}d_{i}d_{j}}\prod_{i=1}^{m}\frac{\mu^{n_{i}d_{i}}a^{a_{i}d_{i}}q^{l_{i}d_{i}}(-t)^{t_{i}d_{i}}}{(q;q)_{d_{i}}},
\end{equation}
where $C$ is a~symmetric $m\times m$ matrix and $n_{i}$, $a_{i}$,
$l_{i}$, $t_{i}$ are fixed integers. In consequence, there exist
a~quiver $Q$ which adjacency matrix is equal to $C$ and motivic
generating series 
\begin{equation}
F_{Q}(\boldsymbol{x},q)=\sum_{d_{1},\ldots,d_{m}\geq0}(-q^{1/2})^{\sum_{i,j=1}^{m}C_{ij}d_{i}d_{j}}\prod_{i=1}^{m}\frac{x_{i}^{d_{i}}}{(q;q)_{d_{i}}}
\end{equation}
reduces to the~GM invariant after the change of variables $x_{i}=\mu^{n_{i}}a^{a_{i}}q^{l_{i}}(-t)^{t_{i}}$:
\begin{equation}
F_{K}(\mu,a,q,t)=\left.F_{Q}(\boldsymbol{x},q)\right|_{x_{i}=\mu^{n_{i}}a^{a_{i}}q^{l_{i}}(-t)^{t_{i}}}.\label{eq:GM-Q correspondence t-deformed}
\end{equation}

\end{conj}

Let us look at examples, starting from the~unknot complement (in
unreduced normalisation). Basing on \cite{GGKPS20xx}, we have

\begin{equation}
F_{0_{1}}(\mu,a,q,t)=\frac{(\mu q;q)_{\infty}}{(-\mu at^{3};q)_{\infty}}.
\end{equation}
We immediately see that the~quiver is the~same as in (\ref{eq:Unknot quiver}),
but the change of variables is given by
\begin{equation}
x_{1}=\mu q^{1/2},\qquad x_{2}=-\mu at^{3}.
\end{equation}

For all $T^{(2,2p+1)}$ torus knots (including trefoil and cinquefoil)
the situation is similar. The~general formula (in reduced normalisation)
\cite{GGKPS20xx}

\begin{equation}
\begin{split}F_{T^{(2,2p+1)}}(\mu,a,q,t)=\sum_{0\leq k_{p}\leq\ldots\leq k_{1}} & \mu^{2(k_{1}+\ldots+k_{p})-k_{1}}q^{(k_{1}+k_{2}+\ldots+k_{p})-\sum_{i=2}^{p}k_{i-1}k_{i}}t^{2(k_{1}+\ldots+k_{p})}\\
\times & \frac{(-aq^{-1}t;q)_{k_{1}}(\mu;q^{-1})_{k_{1}}}{(q;q)_{k_{1}}}\left[\begin{array}{c}
k_{1}\\
k_{2}
\end{array}\right]\cdots\left[\begin{array}{c}
k_{p-1}\\
k_{p}
\end{array}\right]
\end{split}
\end{equation}
leads to the~quiver (\ref{eq:torus knots quiver}-\ref{eq:torus knots quiver elements}),
the only difference with respect to section \ref{subsec:General--torus}
lies in the~change of variables. Now $x_{i}=\mu^{n_{i}}a^{a_{i}}q^{l_{i}}(-t)^{t_{i}}$,
where
\begin{equation}
\begin{split}\sum_{i}t_{i}d_{i}= & 3(d_{1}+d_{3})+2(d_{2}+d_{4})\\
+ & 5(d_{5}+d_{7})+4(d_{6}+d_{8})\\
\vdots\\
+ & (2p+1)(d_{4p-3}+d_{4p-1})+2p(d_{4p-2}+d_{4p}).
\end{split}
\end{equation}
For the trefoil it leads to the~change of variables
\begin{equation}
x_{1}=-\mu^{2}at^{3},\qquad x_{2}=\mu^{2}q^{3/2}t^{2},\qquad x_{3}=-\mu aq^{-1/2}t^{3},\qquad x_{4}=\mu qt^{2}.\label{eq:3_1 quiver change of vars t refined}
\end{equation}
Comparing with the~quiver adjacency matrix (\ref{eq:adjacency matrix trefoil quiver}),
we can see that
\begin{equation}
t_{i}\neq C_{ii},
\end{equation}
in contrary to the~standard knots-quivers correspondence \cite{KRSS1707short,KRSS1707long}.

\subsection{BPS states and 3d $\mathcal{N}=2$ effective theories}

In section \ref{sec:Counting-BPS-states} we constructed knot complement
analogs of LMOV invariants basing on the~DT invariants. Knowing the
form of the~$t$-deformed change of variables $x_{i}=\mu^{n_{i}}a^{a_{i}}q^{l_{i}}(-t)^{t_{i}}$,
we can define the~$t$-deformed knot complement analogs of LMOV invariants:
\begin{equation}
N(\mu,a,q,t)=\sum_{r,i,j,k}N_{r,i,j,k}\mu^{r}a^{i}q^{j}(-t)^{k}=\left.\Omega(\boldsymbol{x},q)\right|_{x_{i}=\mu^{n_{i}}a^{a_{i}}q^{l_{i}}(-t)^{t_{i}}},
\end{equation}
which implies
\begin{equation}
\begin{split}F_{K}(\mu,a,q,t)= & \textrm{Exp}\left[\frac{N(\mu,a,q,t)}{1-q}\right]=\exp\left[\sum_{n,r,i,j}\frac{N_{r,i,j,k}\mu^{rn}a^{in}q^{jn}(-t)^{kn}}{1-q}\right]\\
= & \prod_{r,i,j,k,l}\left(1-\mu^{r}a^{i}q^{j+l}(-t)^{k}\right)^{-N_{r,i,j,k}}=\prod_{r,i,j,k}(\mu^{r}a^{i}q^{j}(-t)^{k};q)_{\infty}^{-N_{r,i,j,k}}.
\end{split}
\label{eq:F_K LMOV invariants t refined}
\end{equation}
Since DT invariants are integer numbers, so are $N_{r,i,j,k}$.

In case of the~unknot complement, the~GM invariant given in terms
of infinite $q$-Pochhammers
\begin{equation}
F_{0_{1}}(\mu,a,q,t)=\frac{(\mu q;q)_{\infty}}{(\mu a(-t)^{3};q)_{\infty}}
\end{equation}
immediately leads to
\begin{equation}
N_{1,0,1,0}=-1,\qquad N_{1,1,0,3}=1.
\end{equation}

From the point of view of 3d $\mathcal{N}=2$ effective theory $T[M_{K}]$,
the~$t$-deformation of the semiclassical limit (\ref{eq:Semiclassical limit F_K})
is given by
\begin{equation}
F_{K}(\mu,a,q,t)\underset{\hbar\rightarrow0}{\rightarrow}\int\prod_{i}\frac{dz_{i}}{z_{i}}\exp\left[\frac{1}{\hbar}\widetilde{\mathcal{W}}_{T[M_{K}]}(\mu,a,t,z_{i})+\mathcal{O}(\hbar^{0})\right]\label{eq:Semiclassical limit F_K t-deformed}
\end{equation}
and can be interpreted as the~introduction of the~global $R$-symmetry
$\textrm{U}(1)_{F}$ with fugacity $(-t)$, associated to rotations
in the directions normal to M5-brane inside $\mathbb{R}^{4}$ (for
more details see \cite{AFGS1203,FGS1205,FGSS1209,EKL1811}). Then
the~moduli space of vacua of $T[M_{K}]$, obtained by the~extremalisation
of $\widetilde{\mathcal{W}}_{T[M_{K}]}$ with respect to~$z_{i}$
and the~introduction of variable $\lambda$ dual to $\mu$, is described
by the~graph of the super-$A$-polynomial~\cite{FGS1205}:
\begin{equation}
\frac{\partial\widetilde{\mathcal{W}}_{T[M_{K}]}^{\textrm{eff}}(\mu,a,t)}{\partial\log\mu}=\log\lambda\qquad\Leftrightarrow\qquad A(\mu,\lambda,a,t)=0.
\end{equation}
$A(\mu,\lambda,a,q,t)$ is the~classical limit of the~quantum super-$A$-polynomial,
which annihilates superpolynomials (see \cite{DGR0505,AFGS1203,FGS1205,FGSS1209}).

\section{Future directions\label{sec:Future-directions}}

We conclude with a~brief description of interesting directions for
future research:
\begin{itemize}
\item It would be desirable to understand the~relation between quivers
found in this paper and in \cite{KRSS1707short,KRSS1707long}. We
could see that for the~unknot and the~unknot complement quivers
are the~same, but the~quiver corresponding to the~trefoil has 3
nodes, whereas the~one associated to the~trefoil complement has
4 nodes. The~core of this issue probably lies in the~relation between
HOMFLY-PT polynomials and GM~invariants. It seems that in general
it should be some transformation analogous to the~Fourier transform,
which for some cases (including $T^{(2,2p+1)}$ torus knots) reduces
to the~substitution $\mu=q^{r}$. Investigating the~relation between
the~standard and new correspondence could help in finding quivers
for other knot complements. Probably it would also provide a new insight
to the duality web of associated 3d $\mathcal{N}=2$ theories studied
in \cite{EKL1811,EKL1910}.
\item Since GPPV~invariants exhibit peculiar modularity properties and
are related to logarithmic conformal field theories, it would be interesting
to perform a~study of these aspects for GM~invariants in the~context
of the~correspondence with quivers.
\item GM~invariants still lack a~proper mathematical definition and a~proof
that they are ineed topological invariants of knot complements. These
goals seem ambitious, but achieving them would potentially allow to
properly state and prove the~conjectures proposed in this work.
\item One can look for suggestions and constraints on GM~invariants coming
from the~existence of a~corresponding quiver. Such approach applied
to the~case of the~standard knots-quivers correspondence enabled
finding the~formulas for HOMFLY-PT polynomials $P_{K,r}(a,q)$ for
$6_{2}$ and $6_{3}$ knots \cite{KRSS1707long}.
\item The~form and meaning of the~$t$-deformation proposed in \cite{GGKPS20xx}
and applied here are still uncertain. Better understanding of this
aspect is crucial for the~categorification of 3-manifold invariants,
which was the~main motivation of introducing GPPV and GM~invariants.
\end{itemize}

\section*{Acknowledgements}

I would like to thank Tobias Ekholm, Angus Gruen, Sergei Gukov, Pietro
Longhi, Sunghuk Park, and Piotr Su\l{}kowski for insightful discussions.
My work is supported by the Polish Ministry of Science and Higher
Education through its programme Mobility Plus (decision no. 1667/MOB/V/2017/0).

\newpage{}

\bibliographystyle{JHEP}
\bibliography{Quivers}

\end{document}